%% file: main.tex
\documentclass[sigconf, nonacm, 10pt]{acmart}
\renewcommand\footnotetextcopyrightpermission[1]{}
\acmConference{}
\acmBooktitle{}
\acmYear{}
\copyrightyear{}
\usepackage{booktabs} 

\usepackage[ruled, vlined, linesnumbered]{algorithm2e}
\usepackage{amsfonts}
\usepackage{amsmath}

\usepackage{amssymb, bm}
\usepackage{amsthm}
\usepackage{cases}
\usepackage{enumitem}
\usepackage{gensymb}
\usepackage{soul}
\usepackage{mathtools}
\usepackage{tcolorbox}
\usepackage{tikz}
\usepackage[caption=false,font=footnotesize]{subfig}

\renewcommand\footnotetextcopyrightpermission[1]{}

\begin{document}
\thispagestyle{plain}

\setlength{\abovedisplayskip}{3pt}
\setlength{\belowdisplayskip}{3pt}

\makeatletter
\fancyhead[L]{}              
\makeatother

\input{macro}


\title[]{{\namebf}: Exploring and Benchmarking of LDPC Decoding across CPU, GPU, and ASIC Platforms}

\author{Zhenzhou Qi$^{1}$, Yuncheng Yao$^{2}$, Yiming Li$^{1}$, Chung-Hsuan Tung$^{1}$, Junyao Zheng$^{1}$, Danyang Zhuo$^{2}$, Tingjun Chen$^{1}$}
\affiliation{%
  \institution{Departments of Electrical and Computer Engineering$^{1}$ and Computer Science$^{2}$, Duke University}
  \city{}
  \state{}
  \country{}
}

\renewcommand{\shortauthors}{Qi, et al.}



\begin{abstract}
Emerging virtualized radio access networks (vRANs) demand flexible and efficient baseband processing across heterogeneous compute substrates. 
In this paper, we present {\name}, a unified benchmarking framework for evaluating low-density parity-check (LDPC) decoding acceleration across different hardware platforms.
{\name} integrates a comprehensive suite of LDPC decoder implementations, including kernels, APIs, and test vectors for CPUs (FlexRAN), GPUs (Aerial and Sionna-RK), and ASIC (ACC100), and can be readily extended to additional architectures and configurations.
Using {\name}, we systematically characterize how different platforms orchestrate computation--from threading and memory management to data movement and accelerator offload--and quantify the resulting decoding latency under varying Physical layer parameters. 
Our observations reveal distinct trade-offs in parallel efficiency and offload overhead, showing that accelerator gains strongly depend on data-movement and workload granularity. 
Building on these insights, we discuss how cross-platform benchmarking can inform adaptive scheduling and co-design for future heterogeneous vRANs, enabling scalable and energy-efficient baseband processing for NextG wireless systems.
\end{abstract}
%
%
%

%
%

\maketitle

\input{1.introduction}
\input{2.related_works}
\input{4.Exploratory}

\input{5.evaluation}
\input{6.discussion}

\input{7.conclusion}

\begin{acks}
This work was supported in part by NSF grants CNS-2211944, CNS-2330333, CNS-2443137, CNS-2450567, and OAC-2503010.
This work was also supported in part by the Center for Ubiquitous Connectivity (CUbiC), sponsored by Semiconductor Research Corporation (SRC) and Defense Advanced Research Projects Agency (DARPA) under the JUMP 2.0 program.
\end{acks}

\bibliographystyle{ACM-Reference-Format}
\bibliography{reference}

\end{document}

%% file: macro.tex
\definecolor{lightgray}{gray}{0.9}
\definecolor{lightblue}{rgb}{0.9,0.9,1}
\definecolor{LightMagenta}{rgb}{1,0.5,1}
\definecolor{red}{rgb}{1,0,0}
\definecolor{brightgreen}{rgb}{0.4, 1.0, 0.0}

\newcommand\couldremove[1]{{\color{lightgray} #1}}
\newcommand{\remove}[1]{}
\newcommand{\move}[2]{ {\textcolor{Purple}{ \bf --- MOVE #1: --- }} {\textcolor{Orchid}{#2}} }

\newcommand{\hlc}[2][yellow]{ {\sethlcolor{#1} \hl{#2}} }
\newcommand\note[1]{\hlc[SkyBlue]{-- #1 --}} 

\newcommand\mynote[1]{\hlc[yellow]{#1}}
\newcommand\tingjun[1]{\hlc[yellow]{TC: #1}}
\newcommand\zhihui[1]{\hlc[LightMagenta]{ZG: #1}}
\newcommand\tom[1]{\hlc[brightgreen]{TOM: #1}}
\newcommand\andy[1]{\hlc[pink]{Andy: #1}}
\newcommand\change[1]{{\color{blue} {#1}}}

\newcommand{\TODO}[1]{\textcolor{red}{#1}}
\newcommand{\revise}[1]{\textcolor{blue}{#1}}


\newcommand{\myparatight}[1]{\vspace{0.5ex}\noindent\textbf{#1~~}}

\newcommand{\cmark}{\ding{51}}%
\newcommand{\xmark}{\ding{55}}%
\newcommand{\greencheck}{\color[HTML]{3C8031}{\cmark}}
\newcommand{\redcross}{\color[HTML]{ED1B23}{\xmark}}
\newcommand{\greenno}{\color[HTML]{3C8031}{\textbf{No}}}
\newcommand{\redyes}{\color[HTML]{ED1B23}{\textbf{Yes}}}
\newcommand{\greenlow}{\color[HTML]{3C8031}{\textbf{Low}}}
\newcommand{\redhigh}{\color[HTML]{ED1B23}{\textbf{High}}}

\newcommand*\circledwhite[1]{\tikz[baseline=(char.base)]{
            \node[shape=circle,draw,inner sep=0.6pt] (char) {\scriptsize{#1}};}}

\newcommand*\circled[1]{\tikz[baseline=(char.base)]{
            \node[shape=circle,draw,fill=black,text=white,inner sep=0.6pt] (char) {\scriptsize{#1}};}}

\newcommand{\name}{{\sf DecodeX}}
\newcommand{\namebf}{{\sf\textbf{DecodeX}}}
\newcommand{\nameacc}{{\sf DecodeACC}}
\newcommand{\nameaccbf}{{\sf\textbf{DecodeACC}}}
\newcommand{\namecpu}{{\sf DecodeCPU}}
\newcommand{\namecpubf}{{\sf\textbf{DecodeCPU}}}
\newcommand{\namegpu}{{\sf DecodeGPU}}
\newcommand{\namegpubf}{{\sf\textbf{DecodeGPU}}}
\newcommand{\namegpujetson}{{\sf DecodeGPU-Jetson}}
\newcommand{\namegpujetsonbf}{{\sf\textbf{DecodeGPU-Jetson}}}

\newcommand{\savannah}{Savannah}
\newcommand{\savannahsc}{Savannah-sc}
\newcommand{\savannahmc}{Savannah-mc}
\newcommand{\agora}{Agora}
\newcommand{\agorabf}{\textbf{Agora}}

\newcommand{\armavec}{\sf Savannah-mc (arma-vec)}
\newcommand{\armacube}{\sf Savannah-mc (arma-cube)}

\newcommand{\specialcell}[2][c]{%
  \begin{tabular}[#1]{@{}c@{}}#2\end{tabular}}

\newcommand{\iu}{{j}}

\newcommand{\littlesum}{\mathop{\textstyle\sum}}
\newcommand{\littleint}{\mathop{\textstyle\int}}

\newcommand{\siso}{SISO}
\newcommand{\mimoTwoByTwo}{2$\times$2}
\newcommand{\mimoFourByFour}{4$\times$4}

\newcommand{\myCodeShort}[1]{\texttt{\small{#1}}}

\newcommand{\bbdev}{\textsf{bbdev}}

\newcommand{\scs}{{\Delta f}}
\newcommand{\scNum}{N_{\textrm{sc}}}
\newcommand{\sampRate}{F_{\textrm{s}}}
\newcommand{\fftSize}{N_{\textrm{FFT}}}
\newcommand{\chMat}{{\textbf{H}}}
\newcommand{\chVec}{{\textbf{h}}}
\newcommand{\precodeMat}{{\textbf{P}}}

\newcommand{\usec}{$\upmu$s} 
\newcommand{\msec}{ms}       

\newcommand{\fft}{\textsf{fft}}
\newcommand{\ifft}{\textsf{ifft}}
\newcommand{\csi}{\textsf{csi}}
\newcommand{\precode}{\textsf{precode}}
\newcommand{\encode}{\textsf{enc}}
\newcommand{\decode}{\textsf{dec}}
\newcommand{\modul}{\textsf{modul}}
\newcommand{\demod}{\textsf{demod}}
\newcommand{\equal}{\textsf{equal}}

\newcommand{\tbSize}{T}
\newcommand{\tbCrcSize}{T_{\textrm{crc}}}
\newcommand{\cbSize}{K_{\textrm{cb}}}
\newcommand{\cbNum}{N_{\textrm{cb}}}
\newcommand{\liftingSize}{Z_{c}}
\newcommand{\liftingSizeSet}{\mathbf{\Theta}}
\newcommand{\fillerBitNum}{N_{\textrm{filler}}}

\newcommand{\codeRate}{R}

\newcommand{\throughput}{Tp}
\newcommand{\codingTime}{t}
\newcommand{\informationBits}{K'}

\newcommand{\latency}{\mathcal{L}}
\newcommand{\power}{P}

\newcommand{\cellConfigMIMO}{M}
\newcommand{\cellConfigBW}{B}
\newcommand{\cellConfigTL}{\rho_{t}}
\newcommand{\cellConfigTBW}{\rho_{f}}
\newcommand{\cellConfigMCS}{\chi}
\newcommand{\cellConfigVar}{\bm{\upsigma}}
\newcommand{\cellConfig}[5]{\bm{\upsigma}({#1}, {#2}, {#3}, {#4}, {#5})}
\newcommand{\cellProfile}{\mathcal{P}}

\newcommand{\sysConfigVar}{\bm{\uptheta}}
\newcommand{\sysConfig}[4]{\bm{\uptheta}({#2}, {#3} | {#1}, {#4})}
\newcommand{\sysConfigOpt}{\bm{\uptheta}^{\star}}

\newcommand{\setSysConfigVar}{\bm{\upTheta}}
\newcommand{\setSysConfig}[2]{\bm{\upTheta}({#1}, {#2})}
\newcommand{\setSysConfigReduced}[2]{\bm{\upTheta}'({#1}, {#2})}

\newcommand{\sysConfigCellIdx}[5]{\bm{\uptheta}_{#1}({#3}, {#4} | {#2}, {#5})}

\newcommand{\setSize}[1]{|{#1}|}

\newcommand{\numCoreMax}{C_{\textrm{max}}}
\newcommand{\setCoreTotal}{\mathcal{C}}
\newcommand{\numCoreTotal}{C}
\newcommand{\setCoreTotalCellIdx}[1]{\mathcal{C}_{#1}}
\newcommand{\numCoreTotalCellIdx}[1]{C_{#1}}

\newcommand{\setCoreDsp}{\mathcal{C}^{\textrm{dsp}}}
\newcommand{\numCoreDsp}{C^{\textrm{dsp}}}
\newcommand{\setCoreDspCellIdx}[1]{\mathcal{C}_{#1}^{\textrm{dsp}}}
\newcommand{\numCoreDspCellIdx}[1]{C_{#1}^{\textrm{dsp}}}

\newcommand{\setCoreAcc}{\mathcal{C}^{\textrm{acc}}}
\newcommand{\numCoreAcc}{C^{\textrm{acc}}}
\newcommand{\setCoreAccCellIdx}[1]{\mathcal{C}_{#1}^{\textrm{acc}}}
\newcommand{\numCoreAccCellIdx}[1]{C_{#1}^{\textrm{acc}}}

\newcommand{\numVfAccMax}{V_{\textrm{max}}}
\newcommand{\setVfAcc}{\mathcal{V}}
\newcommand{\numVfAcc}{V}
\newcommand{\setVfAccCellIdx}[1]{\mathcal{V}_{#1}}
\newcommand{\numVfAccCellIdx}[1]{V_{#1}}

\newcommand{\costRatioDspAcc}{\alpha}

\newcommand{\myAbs}[1]{\left|{#1}\right|}
\newcommand{\myAng}[1]{\angle{#1}}
\newcommand{\myConjugate}[1]{{#1}^{*}}
\newcommand{\myTranspose}[1]{{#1}^{\top}}
\newcommand{\myHermitian}[1]{{#1}^{H}}
\newcommand{\myIsFunc}[1]{\mathbf{1}\{#1\}}

\newcommand{\AoD}{\phi}
\newcommand{\AoDVec}{\bm{\upphi}}
\newcommand{\AoDbf}{\boldsymbol\phi}
\newcommand{\AoDDirectional}{\Phi}
\newcommand{\az}{\phi}
\newcommand{\azVec}{\bm{\upphi}}
\newcommand{\azVecUE}{\bm{\upphi}_{\textrm{UE}}}
\newcommand{\azbf}{\boldsymbol\phi}
\newcommand{\el}{\psi}
\newcommand{\elbf}{\boldsymbol\psi}

\newcommand{\ElemComp}{w}
\newcommand{\ElemCompbf}{\mathbf{w}}
\newcommand{\ElemCompNew}{w^\prime}
\newcommand{\ElemCompNewbf}{\mathbf{w}^\prime}
\newcommand{\ElemAmp}{A}
\newcommand{\ElemAmpbf}{\mathbf{A}}
\newcommand{\ElemPhase}{\theta}
\newcommand{\ElemPhasebf}{\boldsymbol\theta}
\newcommand{\steer}{s}
\newcommand{\steerVec}{\mathbf{s}}
\newcommand{\steermat}{\mathbf{S}}
\newcommand{\beamPattern}{BP}

\newcommand{\bw}{B}
\newcommand{\carrierFreq}{f_{c}}
\newcommand{\carrierWave}{\lambda}

\newcommand{\csiMat}{\mathbf{H}}

\newcommand{\ASA}[2]{\textrm{ASA}({#1},{#2})}
\newcommand{\antNum}{N}
\newcommand{\antIdx}{n}
\newcommand{\antDist}{d}
\newcommand{\subarrayNum}{M}
\newcommand{\subarraySet}{\mathcal{M}}
\newcommand{\subarrayIdx}{m}
\newcommand{\subarrayAntNum}{N_{s}}
\newcommand{\subarrayAntIdx}{n}
\newcommand{\subarrayAntDist}{d}

\newcommand{\setSubarray}{\mathcal{A}}
\newcommand{\subarrayAlloc}{a}
\newcommand{\subarrayAllocVec}{\mathbf{a}}
\newcommand{\subarrayAllocMat}{\mathbf{A}}
\newcommand{\subarrayAllocSet}{\mathbb{A}}

\newcommand{\bfWeight}{w}
\newcommand{\bfWeightVec}{\mathbf{w}}
\newcommand{\bfAmp}{A}
\newcommand{\bfAmpVec}{\mathbf{A}}
\newcommand{\bfPhase}{\theta}
\newcommand{\bfPhaseVec}{\boldsymbol{\theta}}
\newcommand{\bfGain}{g}
\newcommand{\bfGainSig}[1]{g^{\textrm{sig}}_{#1}}
\newcommand{\bfGainInt}[2]{g^{\textrm{int}}_{{#1}\rightarrow{#2}}}


\newcommand{\userNum}{U}
\newcommand{\userIdx}{u}
\newcommand{\userSet}{\mathcal{U}}

\newcommand{\userNumSub}{K}

\newcommand{\cellFeasibleRF}[1]{\mathcal{F}_{\mathrm{RAF}}\!\left(#1\right)}
\newcommand{\cellProfileRF}[1]{\mathcal{P}_{\mathrm{RAF}}\!\left(#1\right)}

\newcommand{\userSelected}{k}
\newcommand{\userSelectedNum}{K}
\newcommand{\userSelectedSet}{\mathcal{K}}

\newcommand{\userAngle}{\phi}
\newcommand{\userWeight}{\alpha}

\newcommand{\baseSNR}{\gamma}
\newcommand{\SNR}{\mathsf{SNR}}
\newcommand{\SNRMax}{\mathsf{SNR}^{\textrm{max}}}
\newcommand{\SINR}{\mathsf{SINR}}
\newcommand{\SINRMax}{\mathsf{SINR}^{\textrm{max}}}
\newcommand{\Capacity}{T}
\newcommand{\Rate}{R}
\newcommand{\RateMax}{\Rate^{\textrm{max}}}
\newcommand{\RateAvg}{\widebar{\Rate}}
\newcommand{\CapacityMax}{\Tilde{T}}
\newcommand{\suppress}{\alpha}

\newcommand{\RateHist}{\widebar{\Rate}}

\newcommand{\past}{p}
\newcommand{\decay}{\beta}

\newcommand{\RateMean}{\Bar{R}}
\newcommand{\JFI}{\mathsf{JFI}}

\newenvironment{spmatrix}[1]
 {\def\mysubscript{#1}\mathop\bgroup\begin{bmatrix}}
 {\end{bmatrix}\egroup_{\textstyle\mathstrut\mysubscript}}

%% file: 1.introduction.tex
\section{Introduction}
\label{sec:intro}


Virtualized radio access networks (vRANs) are redefining how next-generation cellular systems deploy baseband processing~\cite{qi2024savannah, qi2023programmable, qi2025nexus, schiavo2024cloudric, gao2024mambas, qi2024savannah2}. 
By decoupling Layer-1 (L1) functions from dedicated hardware and running them as software on general-purpose or accelerated platforms, vRANs enable flexible orchestration and cloud-native elasticity. 
However, this shift also exposes severe computational and latency challenges, particularly for the Physical layer forward error correction (FEC) blocks such as LDPC decoding, which dominate the processing time in the 5G uplink and downlink chains.
Recent advances in heterogeneous computing have introduced multiple candidate substrates for vRAN baseband processing acceleration: specialized software library on multicore CPUs (e.g., Intel FlexRAN), GPU-based solutions from both open-source (e.g., OAI CUDA LDPC) and vendor-optimized (NVIDIA Aerial, Sionna-RK) frameworks, and ASICs (e.g., Intel ACC100 eASIC).
Despite their growing availability, comprehensive benchmarking and system-level understanding of how these platforms execute LDPC decoding, interact with memory hierarchies, and trade off latency versus flexibility remains largely unexplored.

\begin{figure}[!t]
    \centering
    \includegraphics[width=0.99\columnwidth]{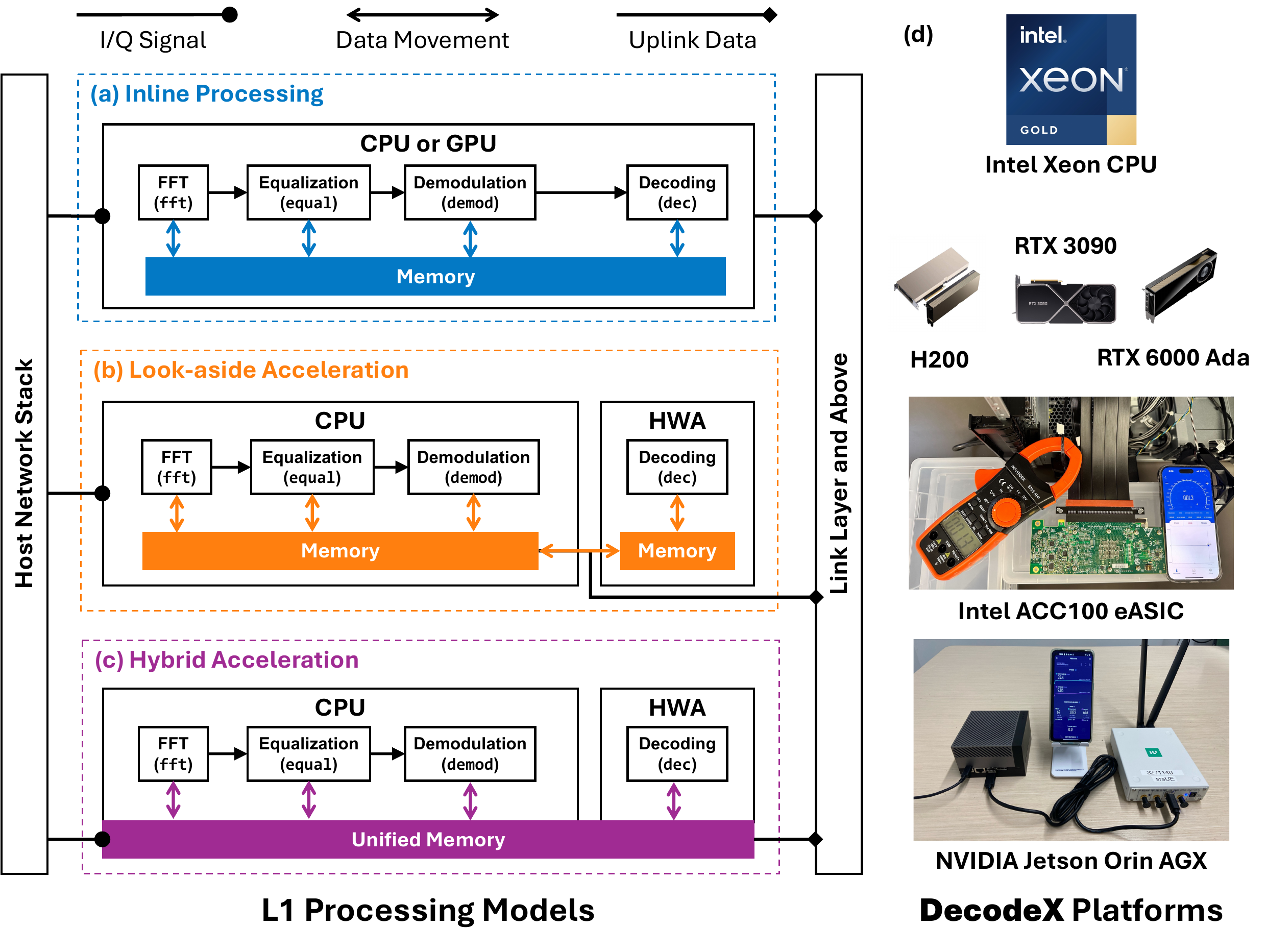}
    \vspace{-3mm}
    \caption{\small
    (a)--(c) Overview of L1 processing models with different hardware accelerators (HWAs):
    Inline Processing, Lookaside Acceleration, and Hybrid Acceleration. 
    (d) Representative {\namebf} implementations of these models: 
    DecodeCPU-FlexRAN (Intel Xeon CPU), 
    DecodeGPU-Aerial (NVIDIA H200, RTX 3090, and RTX 6000 Ada GPUs), 
    DecodeASIC-ACC100 (Intel eASIC), and 
    DecodeGPU-SionnaRK (NVIDIA Jetson Orin AGX).
    }
    \label{fig: compute_flow}
    \vspace{-3mm}
\end{figure}


In this paper, we present {\name}, a cross-platform exploration of LDPC decoding acceleration across CPU, GPU, and ASIC platforms. 
As a comprehensive benchmarking framework, {\name} allows for unified evaluation across different kernel implementations under a consistent interface, each named {\texttt{\small Decode\{HW\}-\{Impl\}}. 
We examine representative implementations of {\name}, including
DecodeCPU-FlexRAN,
DecodeGPU-Aerial,
DecodeGPU-SionnaRK, and
DecodeASIC-ACC100, which collectively capture the diversity of software and hardware acceleration models in modern vRAN deployments. 
{\name} is readily extensible to additional configurations such as 
DecodeCPU-OAI and
DecodeGPU-BBDev, making it a versatile tool for future benchmarking and cross-platform analysis.
We perform a detailed analysis of how each system organizes its decoding pipeline: from threading and SIMD parallelization on CPUs, offload management and queue-based scheduling in DPDK BBDev, to CUDA kernel launches and memory transfers on GPUs. 
Extensive profiling results quantify LDPC decoding latency as a function of modulation and coding scheme (MCS), signal-to-noise ratio (SNR), and physical resource block (PRB) configuration across different {\name} implementations. 
Our measurements reveal that accelerator performance is influenced not only by raw compute throughput but also by data-movement efficiency and task granularity.
To further improve GPU utilization and latency efficiency, we extended NVIDIA's PyAerial framework by introducing a parallel LDPC processing interface, enabling simultaneous decoding of multiple transport blocks across CUDA streams.

Overall, {\name} provides a comparative, system-level characterization of LDPC decoding across heterogeneous vRAN systems. 
Beyond benchmarking, it offers actionable insights for future vRAN baseband and compute co-design, where dynamic resource allocation across CPUs, GPUs, and ASICs will be essential to balancing latency, energy efficiency, and flexibility. 
\emph{To the best of our knowledge, this work presents the first comprehensive study of LDPC decoding performance across different platforms, supported by a complete suite of unit tests and benchmarking scripts open-sourced at~\cite{DecodeX-github}.}


%% file: 2.related_works.tex
\section{Related Work and Background}
\label{sec:related}


The evolution of vRANs has driven extensive research on leveraging heterogeneous computing platforms to accelerate 5G L1 workloads. 
Prior works such as CloudRIC~\cite{schiavo2024cloudric} and Savannah~\cite{qi2024savannah} demonstrate that general-purpose CPUs can execute full 5G baseband pipelines efficiently, while accelerator-based designs explore offloading compute-intensive kernels, such as forward error correction (FEC), to specialized hardware. 
However, these solutions often face challenges in balancing latency, data-movement overhead, and resource sharing across heterogeneous devices, motivating the need for a systematic cross-platform analysis.

\myparatight{5G L1 and LDPC decoding.}
In 5G NR, each slot consists of 14 OFDM symbols~\cite{3gpp38.213}, and depending on the MCS and bandwidth, multiple transport blocks (TBs) may be transmitted per frame. 
A TB corresponds to a data unit scheduled by the MAC layer and may span one or more slots. 
For data channels, 5G NR adopts LDPC codes based on two standardized base graphs (BG1 and BG2)~\cite{3gpp38.212}, each optimized for different code rates and lifting sizes. 
These quasi-cyclic codes compactly represent the parity-check matrix through circulant shifts, avoiding storage of large dense matrices. 
At the receiver, rate de-matching reconstructs code blocks (CBs)~\cite{5g_and_beyond} from TBs, followed by iterative belief-propagation decoding across thousands of variable and check nodes.
The LDPC decoding configuration depends on several key parameters. 
The MCS defines the modulation order ($Q_m$) and determines the code rate after rate matching (via puncturing or repetition). 
The \emph{lifting size} ($\liftingSize$) expands the base graph into the full parity-check matrix. 
A single TB may be segmented into multiple CBs, with the number of decoding iterations adapting to channel quality and SNR. 
LDPC decoding typically dominates the L1 processing latency as it involves fine-grained data dependencies, irregular memory access, and iterative message-passing updates.

\myparatight{Heterogeneous acceleration.}
To meet real-time constraints, diverse acceleration strategies have been explored: CPU implementations rely on SIMD vectorization~\cite{flexran-fec-sdk}, eASICs like Intel's ACC100 exploit deeply pipelined and deterministic architectures, and GPUs leverage massive parallelism for high iteration counts. 
While GPUs provide exceptional throughput, discrete devices often follow a \emph{lookaside offload model}, requiring frequent host-device transfers that increase energy and latency overhead. 
Integrated platforms such as NVIDIA Jetson mitigate these penalties through unified memory, but their limited compute capacity restricts scalability for datacenter or multi-cell vRAN deployments.

{\name} differs from prior efforts by providing a unified, cross-platform benchmarking and profiling framework for LDPC decoding across CPU, GPU, and ASIC architectures. 
It systematically dissects each platform's execution pipeline, from threading and queue scheduling to memory movement and kernel launches, quantifying decoding latency across varying MCS, SNR, and PRB allocations. 

%% file: 4.Exploratory.tex
\section{Design of {\namebf}}
\label{sec:explore}

\subsection{DecodeCPU-FlexRAN}
The DecodeCPU-FlexRAN platform represents a CPU-based inline LDPC decoding environment built upon Intel’s FlexRAN framework illustrated in Fig.~\ref{fig: compute_flow}(a). 
To compile and execute the FlexRAN FEC Software Development Kit (SDK)~\cite{flexran-fec-sdk}, several Intel toolchains are required, including the \texttt{\small oneAPI Base Toolkit}~\cite{oneAPI-base} and the \texttt{\small HPC Toolkit}~\cite{oneAPI-HPC}. 
Depending on the operating system kernel version and system environment, either the Intel \texttt{\small icc} (classic) or \texttt{\small icx} (LLVM-based)~\cite{icc} compiler must be used to ensure compatibility with the vectorized kernels in the FEC SDK. 
These compilers enable efficient vectorization through Intel’s Advanced Vector Extensions (AVX), which form the foundation of the high-performance LDPC decoder in FlexRAN.
The FlexRAN LDPC decoder provides multiple architecture-specific implementations optimized for \texttt{\small  AVX2}/\texttt{\small AVX512}. 
Given that our testbed CPUs support the \texttt{\small AVX512F}, we focus on AVX512, which leverages wide-lane SIMD parallelism to accelerate the layered belief-propagation decoding steps. 
Each decoding thread operates on 512-bit vector registers, processing 32 LLR values concurrently per iteration.
%
Parallelism at the task level is straightforward for CPU-based LDPC processing. 
The smallest schedulable unit is a \emph{TB}, and to fully utilize available CPU resources, multiple TBs can be decoded in parallel across independent cores. 
In practice, decoding $N_{\text{tb}}$ transport blocks concurrently maps to $N_{\text{tb}}$ physical CPU cores (or hyperthreads), enabling near-linear scaling until memory bandwidth saturation. 
This design 
serves as a baseline reference for comparison with hardware-accelerated decoding platforms.

\subsection{DecodeGPU-Aerial}
While DecodeCPU-FlexRAN demonstrates the flexibility of CPU-based baseband processing, its performance is fundamentally limited. 
To improve the processing latency, recent research has explored offloading computationally intensive 5G NR L1 tasks, such as LDPC decoding, to GPUs. 
Earlier implementations, such as the CUDA-based LDPC decoder in OAI~\cite{OAI-CUDA-LDPC} follows a \emph{lookaside} model, where decoded data must traverse the host-device boundary for each transport block. 
Although this approach improves raw decoding throughput, frequent PCIe transfers and high GPU power draw introduce considerable latency and energy overheads. 
To address these challenges, NVIDIA's Aerial platform re-architects the entire L1 processing pipeline as a fully GPU-resident data path.
This \emph{inline} acceleration enables Aerial to execute all L1 functions--from FFT and equalization to LDPC decoding--directly within GPU, minimizing data movement and realizing a practical pathway toward GPU-based vRAN deployments.

\begin{algorithm}[!t]
\footnotesize
\DontPrintSemicolon
\SetAlgoLined
\caption{DecodeGPU-Aerial Interface (Sequential vs.\ Parallel LDPC Decoding)}
\label{algo:pyaerial-decodex}

\textbf{Sequential LDPC processing:}\;
\For{$\texttt{tbIdx} \gets 0$ \KwTo $\texttt{nUEs}-1$}{
  // Prepare decoder configuration, allocate soft-output buffers, and define tensor layouts\;
\texttt{cuphyErrorCorrectionLDPCDecode(LLR)}\;
  // Copy decoded bits back to host memory and advance output pointer\;
  \texttt{gridDim = NUM\_CODEWORDS \;
  launch\_kernel\_driver\_api(gridDim, blockDim)}\;
  
}

\smallskip
\textbf{Parallel LDPC processing:}\;
// Perform the same per-UE preparation (configure decoder, allocate memory, set tensor descriptors)\;
\texttt{cuphyErrorCorrectionLDPCTransportBlockDecode(LLRs)}\;
// Gather decoded outputs for all UEs, finalize buffers, and release temporary allocations\;
\texttt{gridDim = get\_total\_num\_codewords(LLRs)}\;
\texttt{launch\_tb\_kernel\_driver\_api(gridDim, blockDim)}\;

\end{algorithm}

NVIDIA Aerial is a cloud-native, GPU-accelerated vRAN software stack designed to execute the full 5G NR baseband pipeline, encompassing L1 and portions of MAC layer (L2) functions directly on the GPU. 
Its core components include:  
(\emph{i})~\textbf{cuPHY}, which implements 5G NR L1 primitives such as FFT/IFFT, channel estimation, MIMO equalization, and LDPC encoding/decoding;  
(\emph{ii})~\textbf{cuMAC}, which provides scheduling and control integration between MAC and L1 layers; and  
(\emph{iii})~\textbf{pyAerial}, a set of Python APIs for rapid prototyping and testing of GPU-based signal-processing pipelines.  
Together, these modules provide a unified runtime, where baseband workloads are expressed as CUDA kernels operating over shared GPU memory.
Aerial adopts an \emph{inline acceleration} model in which baseband data is transferred directly from the NIC or radio front-end into GPU memory (HWA memory) using GPUNetIO via GPUDirect RDMA~\cite{GPUNet}. 
Once in device memory, the entire L1 chain--from OFDM demodulation to LDPC decoding and rate matching--is executed within the GPU context without host intervention. 
This eliminates intermediate PCIe copies and enables microsecond-level kernel scheduling through CUDA streams and graphs. 
Fig.~\ref{fig: compute_flow}(a) illustrates this architecture, where the data path remains fully resident on the GPU, and only decoded transport blocks are returned to the host for higher-layer processing.

\subsection{DecodeASIC-ACC100}


DecodeASIC-ACC100 integrates a hardware LDPC decoder using Intel’s ACC100 eASIC, which is dedicated to FEC acceleration in vRAN deployments and supports multiple functions, including LDPC encoding and decoding for 5G NR. 
This design follows the \emph{lookaside acceleration} model, as illustrated in Fig.~\ref{fig: compute_flow}(b), where data is explicitly transferred between the host CPU and the accelerator through system memory.
Access to the accelerator is managed through the Data Plane Development Kit (DPDK)~\cite{dpdk-bbdev}, an open-source framework for user-space packet and data-plane processing.
As a baseband device, the ACC100~\cite{acc100} exposes a polling-mode driver (PMD) interface through DPDK, enabling low-latency interaction between user-space applications and the hardware accelerator.
In this section, we briefly describe how DecodeASIC-ACC100 interacts with the ACC100 through DPDK’s \texttt{\small rte\_bbdev} APIs, configure LDPC decoding parameters, and performance profiling.

\myparatight{Driver initialization and configuration.}
The ACC100 driver functions form a subset of DPDK’s \texttt{\small rte\_bbdev} API (``baseband device'' abstraction).
The high-level workflow for initializing and executing LDPC decoding via DPDK includes 
device setup, 
operation configuration, and 
runtime execution.  
During device setup, the application detects the accelerator using \texttt{\small rte\_bbdev\_info\_get()} and initializes communication queues via \texttt{\small rte\_bbdev\_setup\_queues()} and \texttt{\small rte\_bbdev\_\allowbreak queue\_configure()}. The device is then activated with \texttt{\small rte\_\allowbreak bbdev\_start()}.  
Prior to execution, input and output buffers are allocated, and operation descriptors are prepared.
For LDPC decoding, each operation structure specifies the buffer pointers and decoding parameters (see \S\ref{sec:related}).
Once operations are configured, they are enqueued to the device through \texttt{\small rte\_bbdev\_enqueue\_ops()}, and completion is monitored via active polling using \texttt{\small rte\_bbdev\_dequeue\_ops()}.
A successful dequeue indicates the completion of decoding for one TB.

\begin{figure}[!t]
    \centering
    \includegraphics[width=0.95\columnwidth]{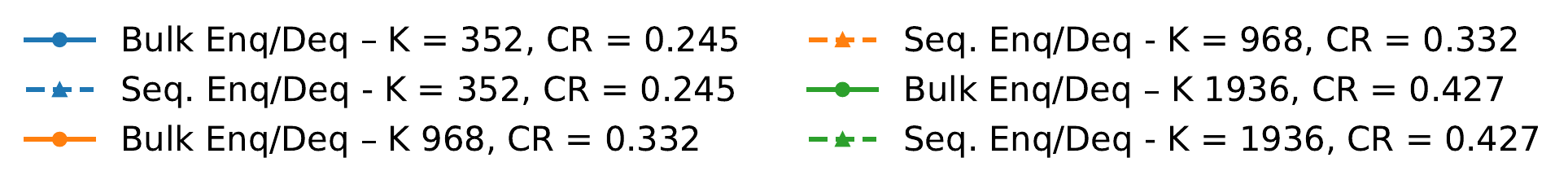}
    
    \vspace{-3mm}
    \subfloat[Intel ACC100]{
        \includegraphics[width=0.45\columnwidth]{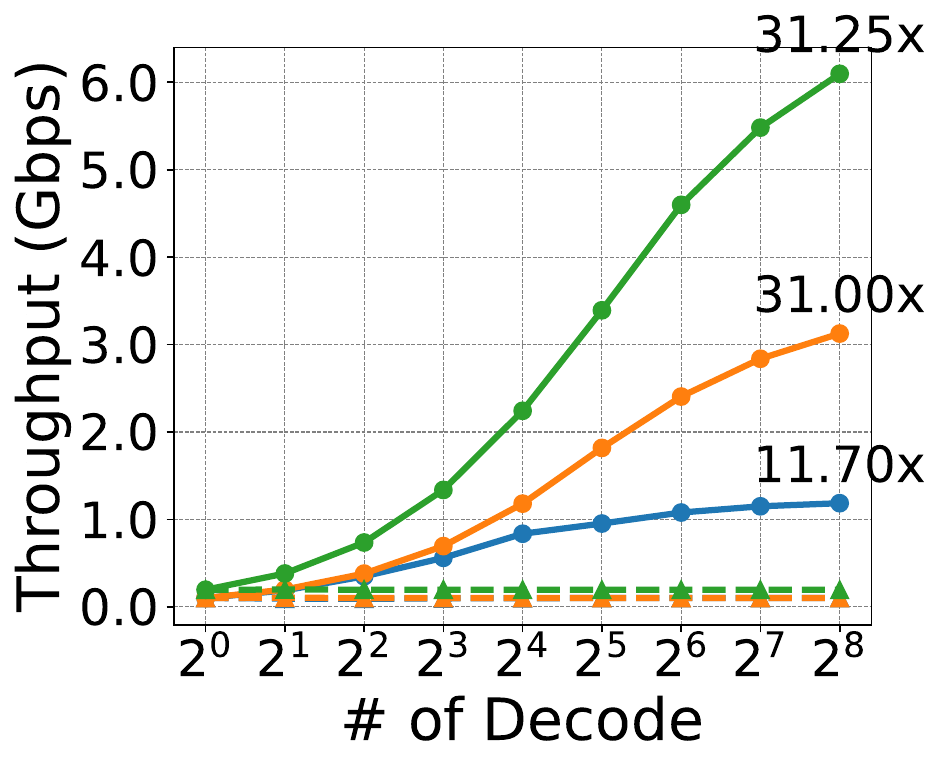}
        \label{fig:intel-tp}}
    \hspace{1mm}
    \vspace{-2mm}
    \subfloat[Silicom ACC100]{
        \includegraphics[width=0.45\columnwidth]{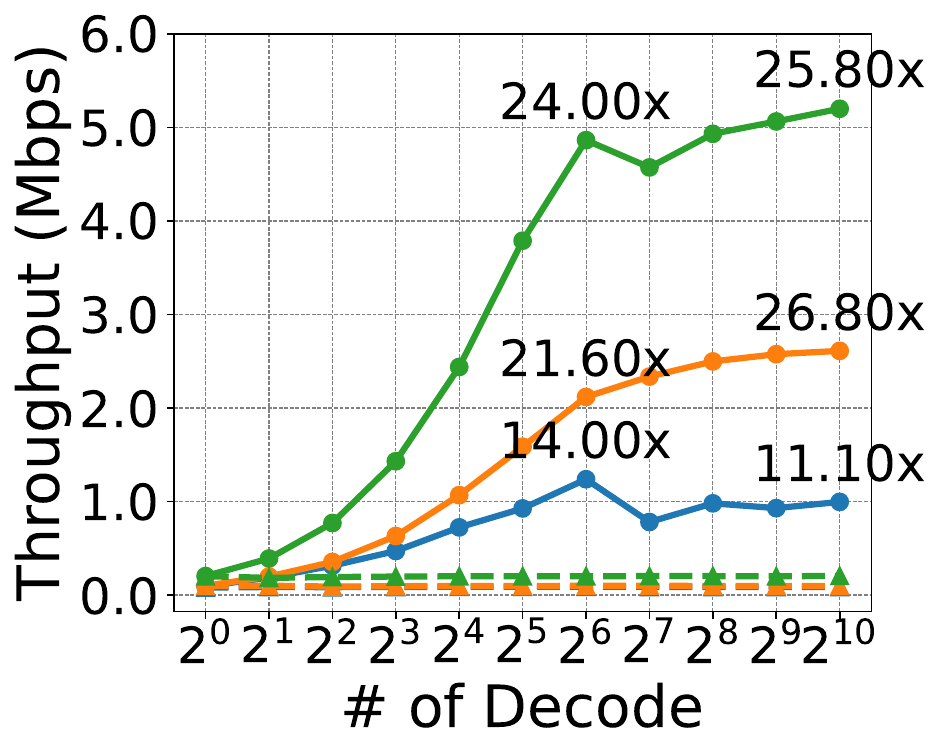}
        \label{fig:silcom-tp}}
    \vspace{-2mm}
    \caption{\small Throughput comparison between sequential and bulk (Algorithm.~\ref{algo:dp-user-selection}) LDPC enqueue and dequeue pipelines on the Intel and Silicom ACC100 platforms.}
    \vspace{-5mm}
    \label{fig: bluk_tp}
\end{figure}

\myparatight{Runtime integration.}
At runtime, each OFDM symbol following {\demod} (i.e., the log-likelihood ratio output, or LLR outputs) is reorganized into an \texttt{\small rte\_bbdev\_op\_data} structure following the BBDev specification, and transferred to the ACC100 device via \texttt{\small rte\_memcpy()}.
To maximize parallelism, DecodeASIC-ACC100 dispatches each demodulated symbol immediately after {\demod} using \texttt{\small rte\_bbdev\_\allowbreak enqueue\_ops()}, and once all enqueue operations are issued, bulk dequeue requests are triggered using \texttt{\small rte\_bbdev\_\allowbreak dequeue\_ops()} to collect decoded results for the entire frame.  
This \emph{parallel enqueue-dequeue strategy} enables continuous data flow, allowing subsequent software tasks (e.g., {\fft} for the next OFDM symbol) to execute concurrently with ongoing hardware decoding.
As shown in Fig.~\ref{fig: bluk_tp}, fully exploiting the ACC100’s intrinsic parallelism through bulk enqueue/dequeue improves decoding throughput by up to {30}$\times$ compared to sequential processing, and the same trend exists across both Intel and Silicom ACC100 variants (Figs.~\ref{fig:intel-tp} and~\ref{fig:silcom-tp}).

\begin{algorithm}[!t]
\footnotesize
\DontPrintSemicolon
\SetAlgoLined
\caption{DecodeASIC-ACC100 Interface}
\label{algo:dp-user-selection}



\BlankLine
\textbf{Run-Time Execution:}

\For{each received symbol}{
  // Read LLR buffers from DSP; enqueue: \\
  \texttt{enq += rte\_bbdev\_enqueue\_ldpc\_dec\_ops(dev, q, ops[enq], 1)}\;

  \If{last UE \textbf{and} last uplink symbol}{
    $deq \gets 0$; $retry \gets 0$\;
    \While{$deq < enq$ \textbf{and} $retry < MAX$}{
      \texttt{deq += rte\_bbdev\_dequeue\_ldpc\_dec\_ops(dev, q, ops\_deq[deq], enq{-}deq)}\;
      $retry{+}{+}$\;
    }
  }
  \Else{ \texttt{continue}\;}
}

\texttt{RTAssert(enq == deq)}\; 

// Compute BLER\;

\end{algorithm}









\begin{figure*}[!t]
    \centering
    \vspace{-3mm}
    \subfloat{
        \includegraphics[width=0.26\textwidth]{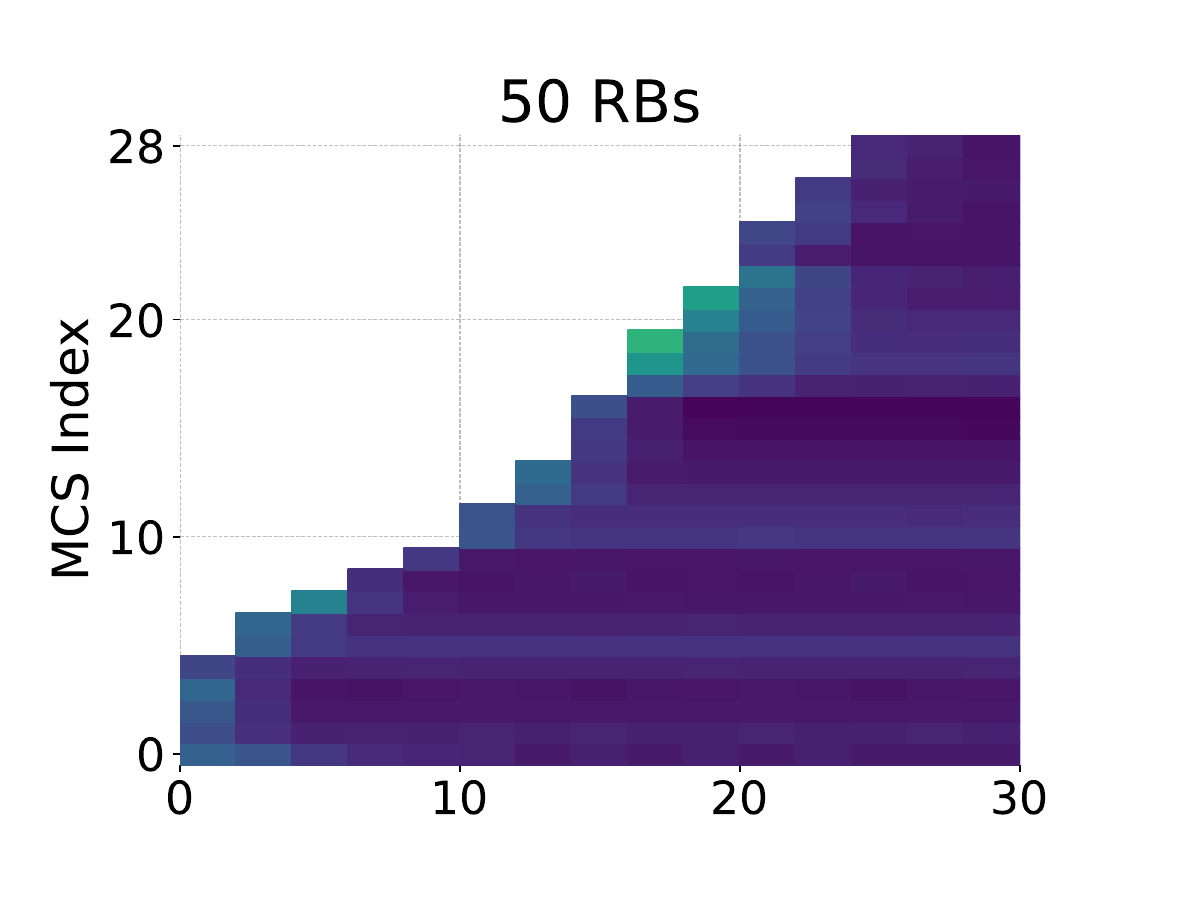}
        \label{fig:cpu-50tb}}
        \hspace{-8mm}
    \subfloat{
        \includegraphics[width=0.26\textwidth]{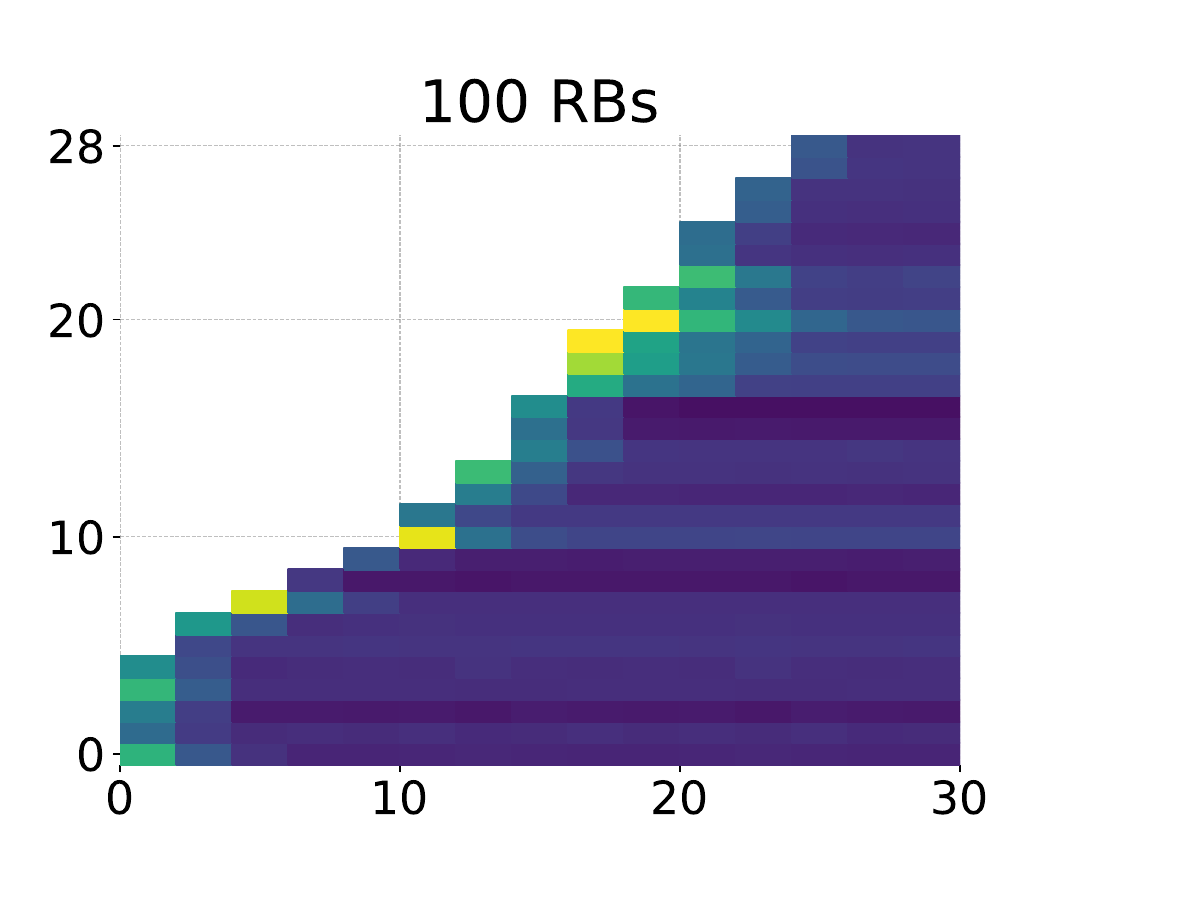}
        \label{fig:cpu-100tb}}
        \hspace{-10mm}
    \subfloat{
        \includegraphics[width=0.26\textwidth]{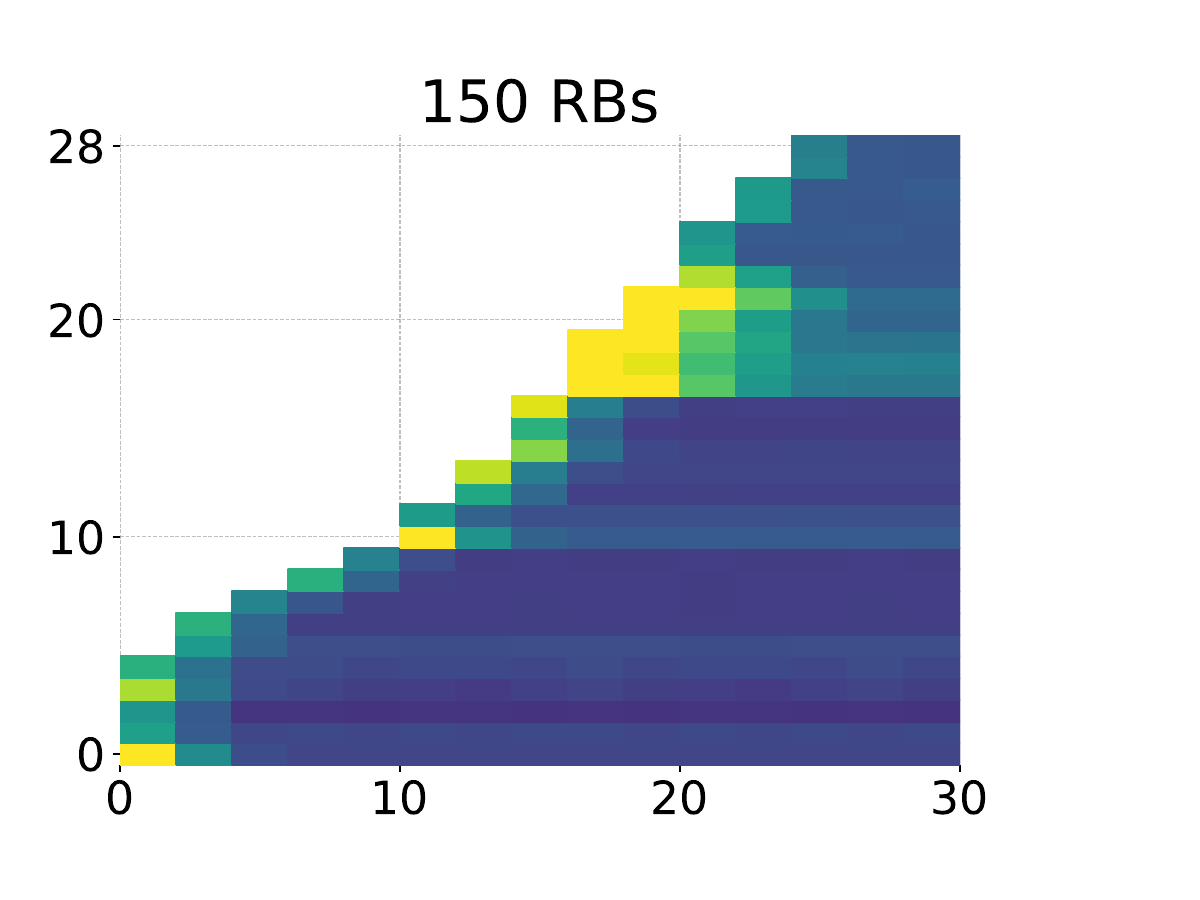}
        \label{fig:cpu-150tb}}
        \hspace{-10mm}
    \subfloat{
        \includegraphics[width=0.26\textwidth]{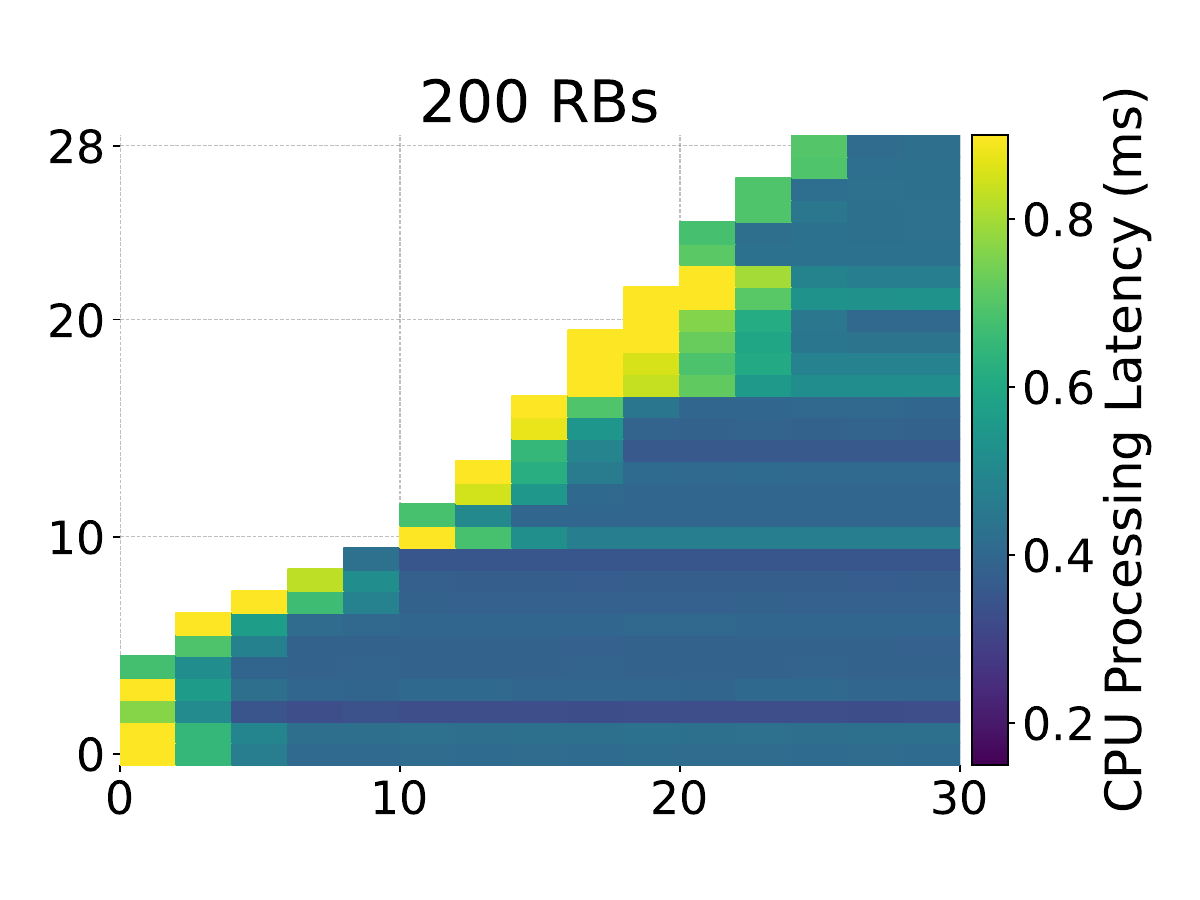}
        \label{fig:cpu-200tb}} \\
        \vspace{-3mm}
    \subfloat{
        \includegraphics[width=0.26\textwidth]{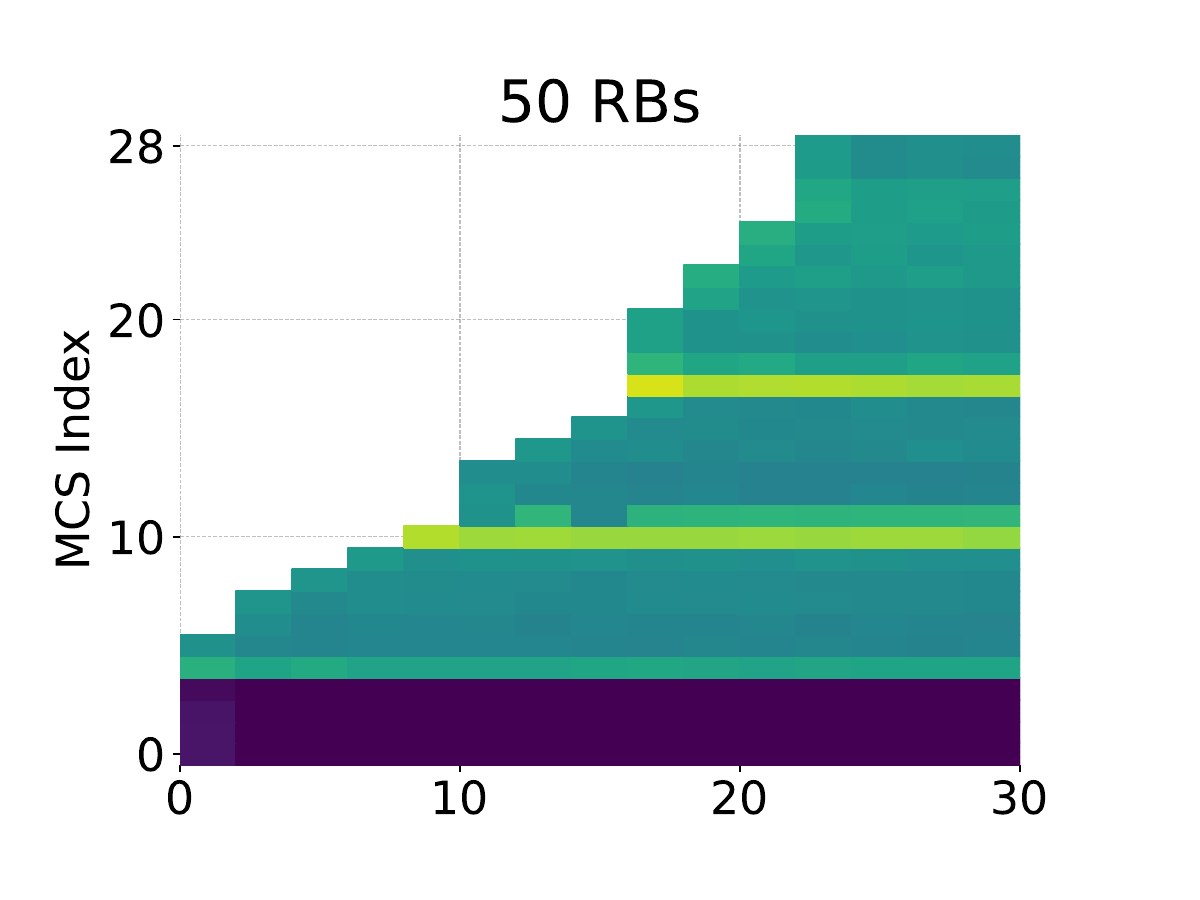}
        \label{fig:acc-50tb}}
        \hspace{-8mm}
    \subfloat{
        \includegraphics[width=0.26\textwidth]{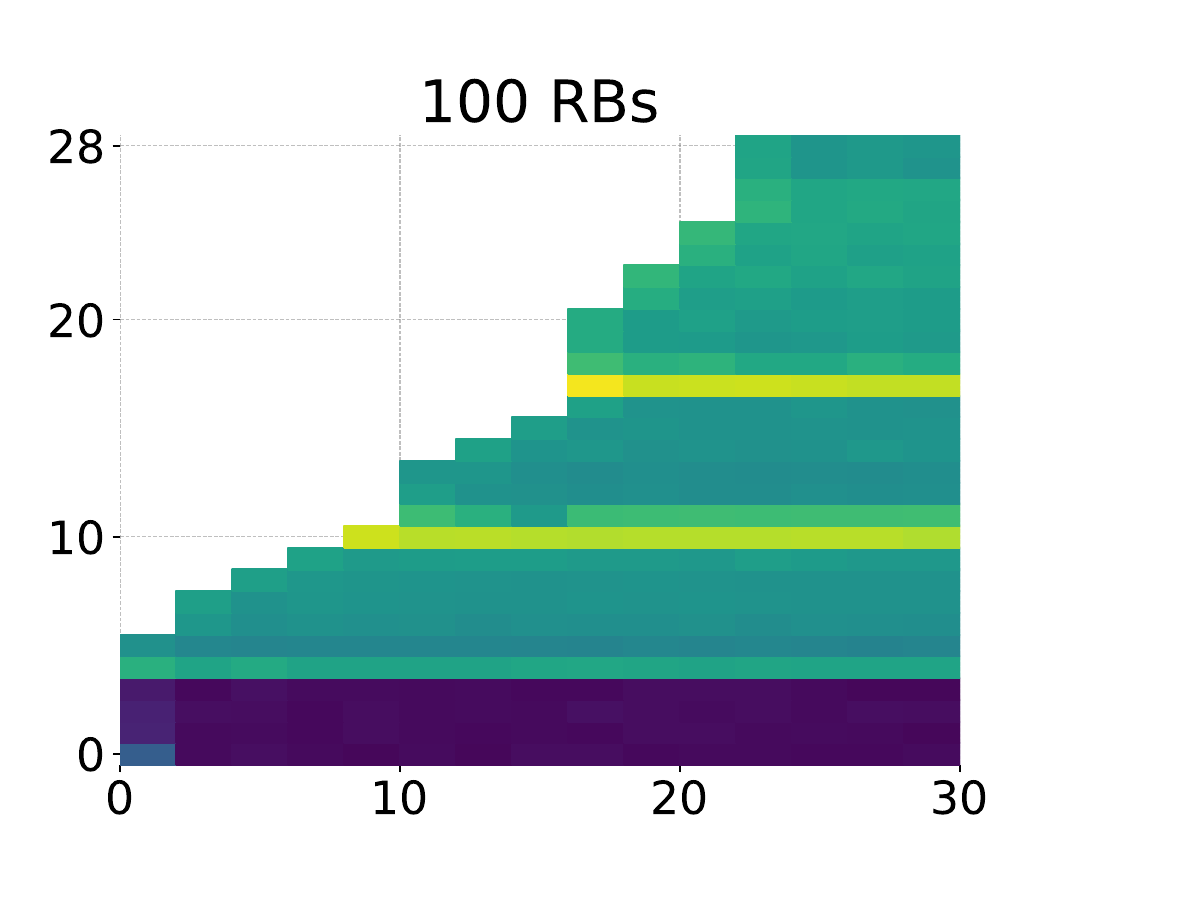}
        \label{fig:acc-100tb}}
        \hspace{-10mm}
    \subfloat{
        \includegraphics[width=0.26\textwidth]{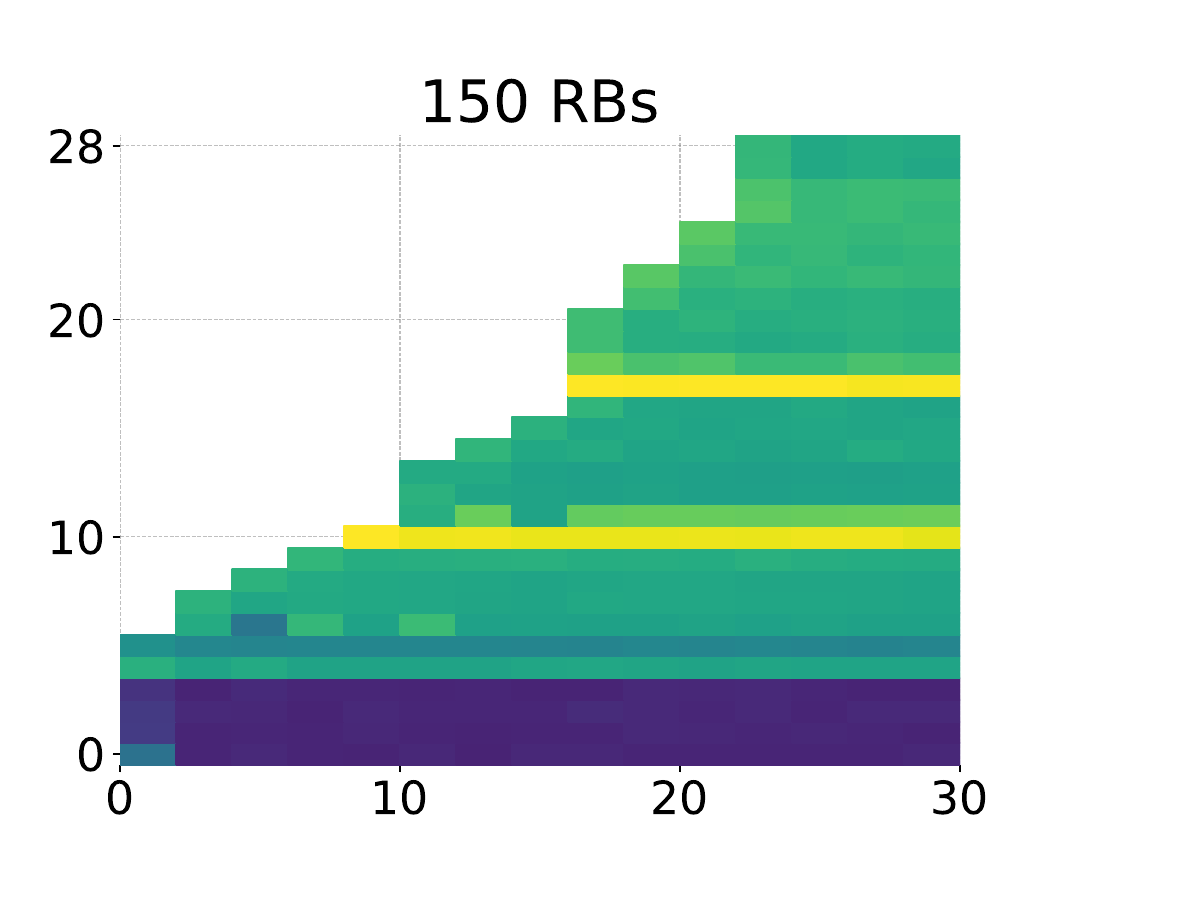}
        \label{fig:acc-150tb}}
        \hspace{-10mm}
    \subfloat{
        \includegraphics[width=0.26\textwidth]{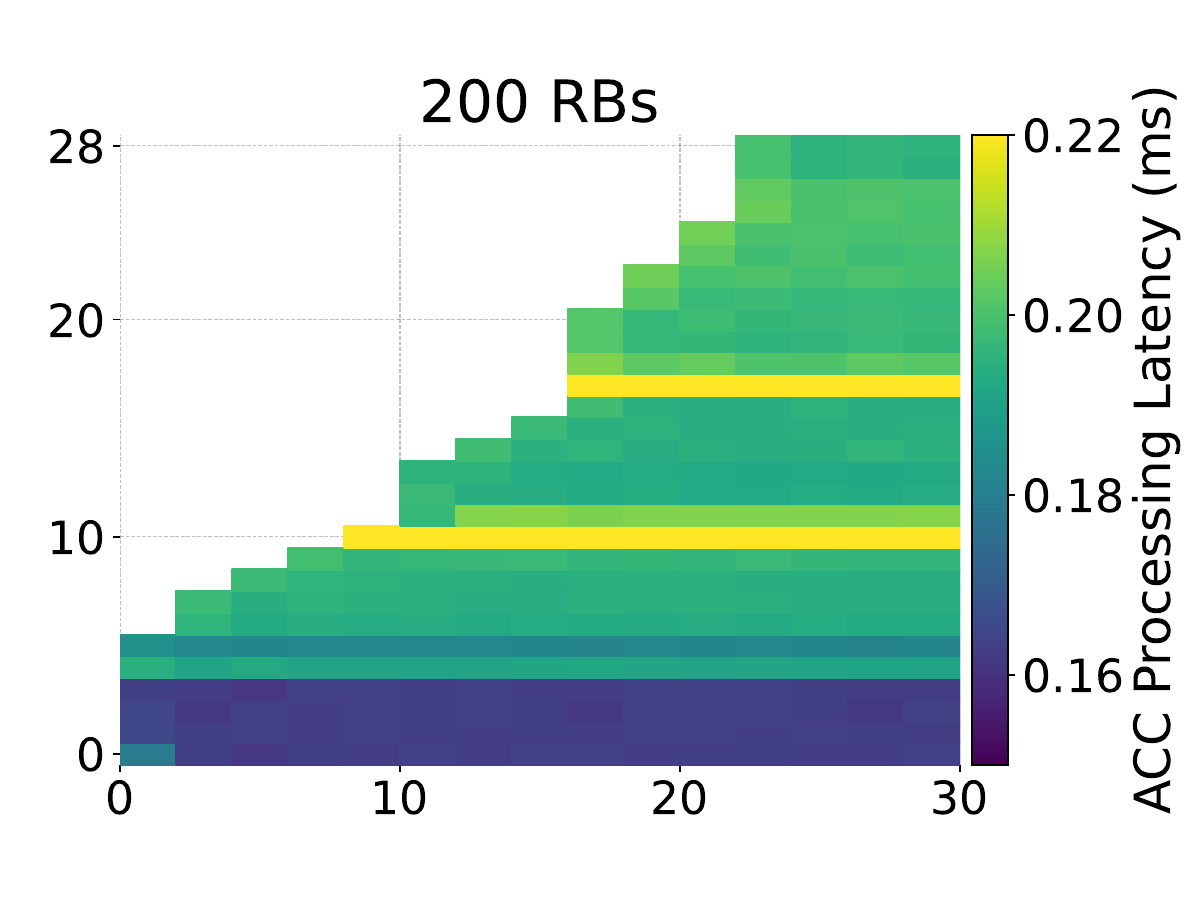}
        \label{fig:acc-200tb}} \\
        \vspace{-3mm}
    \subfloat{
        \includegraphics[width=0.26\textwidth]{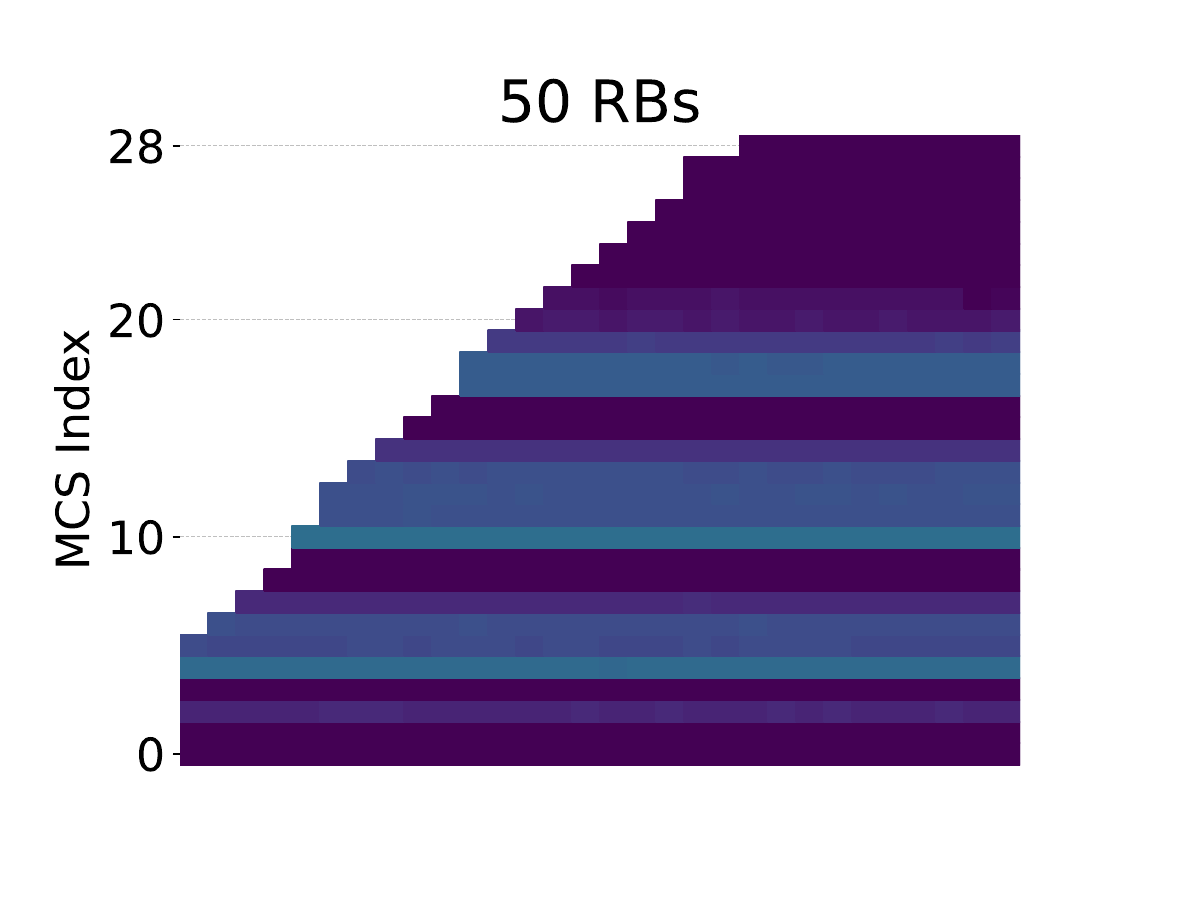}
        \label{fig:gpu-50tb}}
        \hspace{-8mm}
    \subfloat{
        \includegraphics[width=0.26\textwidth]{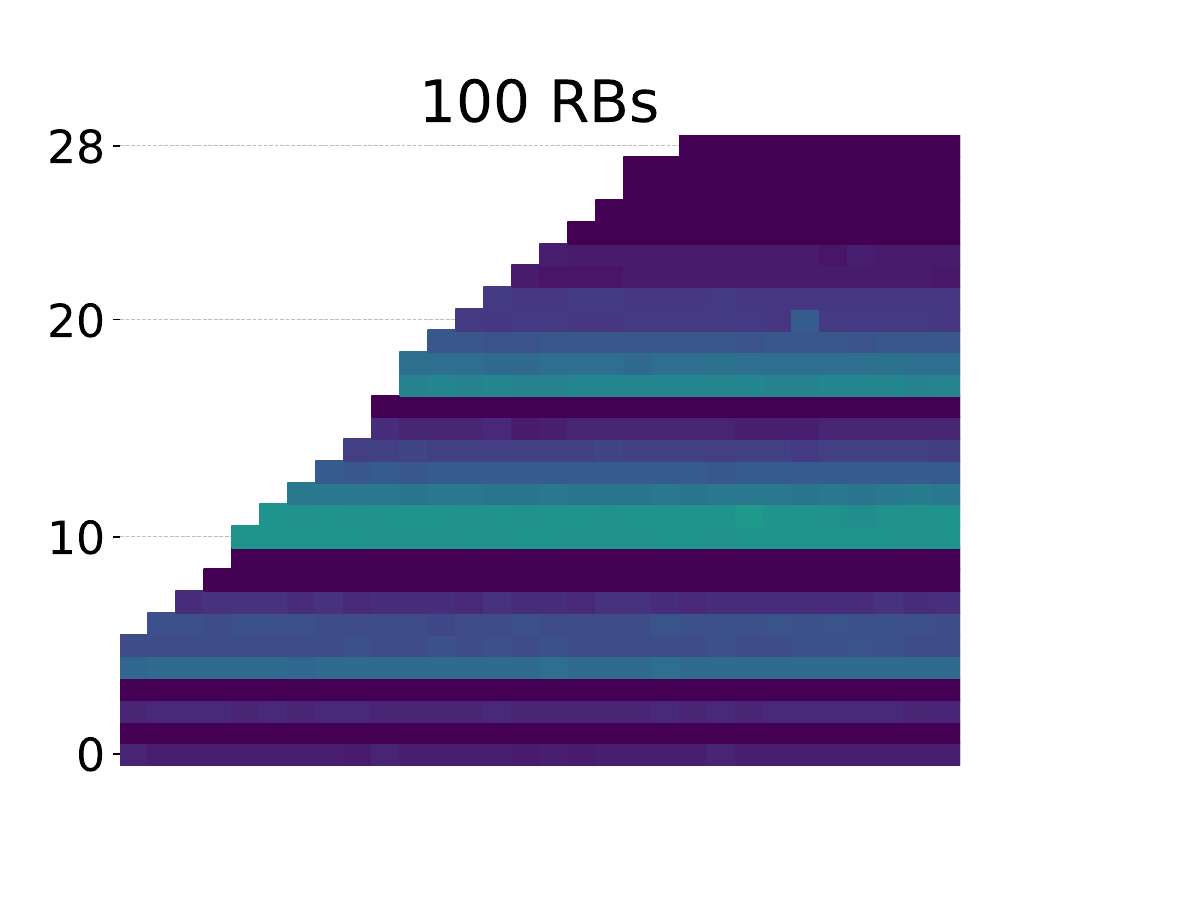}
        \label{fig:gpu-100tb}}
        \hspace{-10mm}
    \subfloat{
        \includegraphics[width=0.26\textwidth]{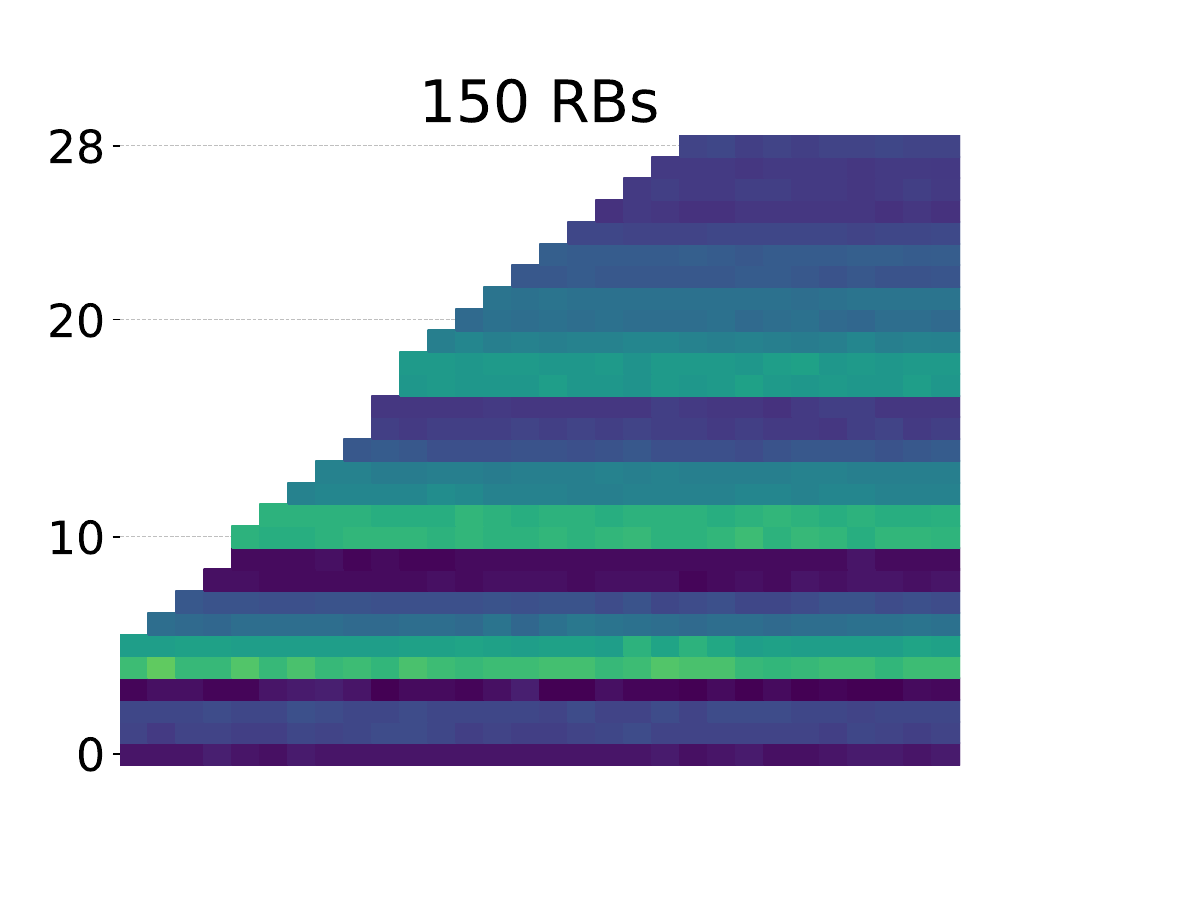}
        \label{fig:gpu-150tb}}
        \hspace{-10mm}
    \subfloat{
        \includegraphics[width=0.26\textwidth]{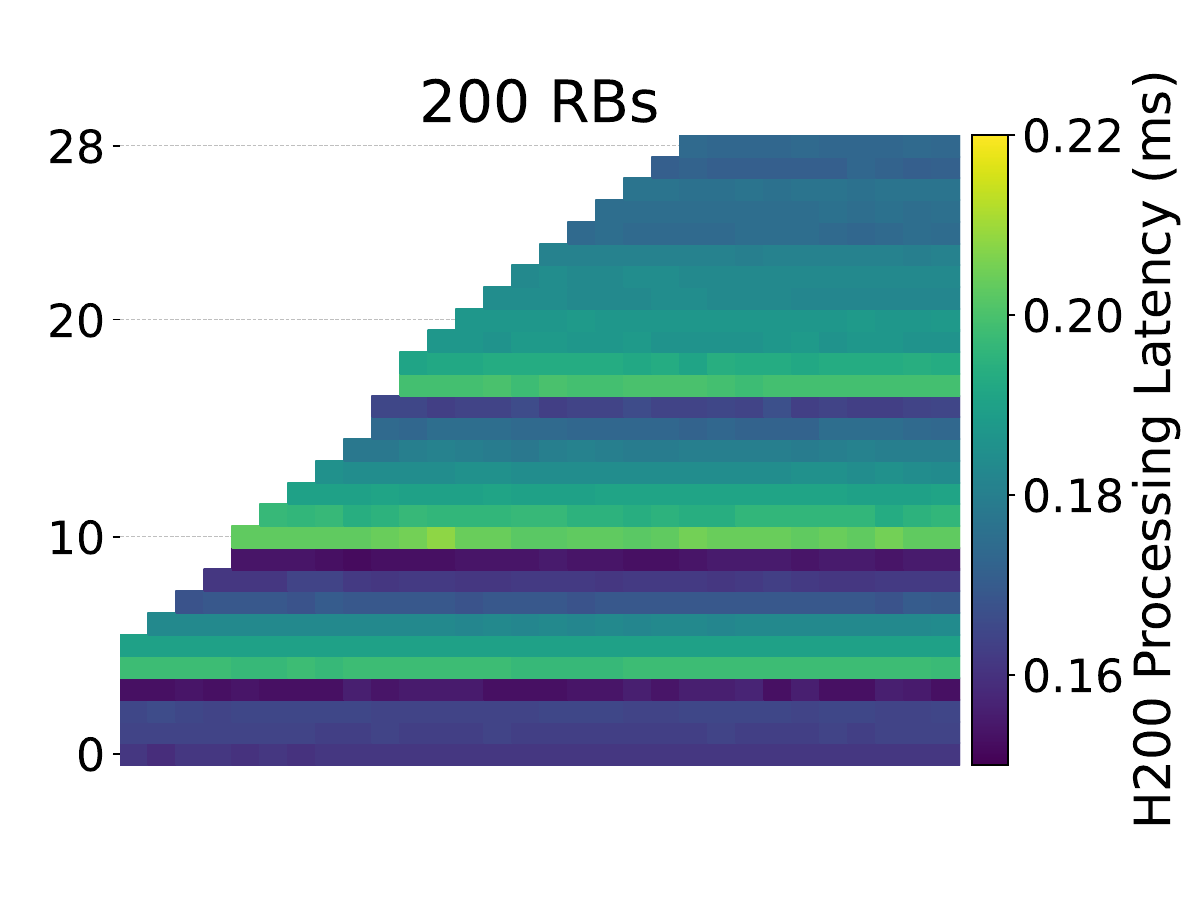}
        \label{fig:gpu-200tb}} \\
        \vspace{-3mm}
    \subfloat{
        \includegraphics[width=0.26\textwidth]{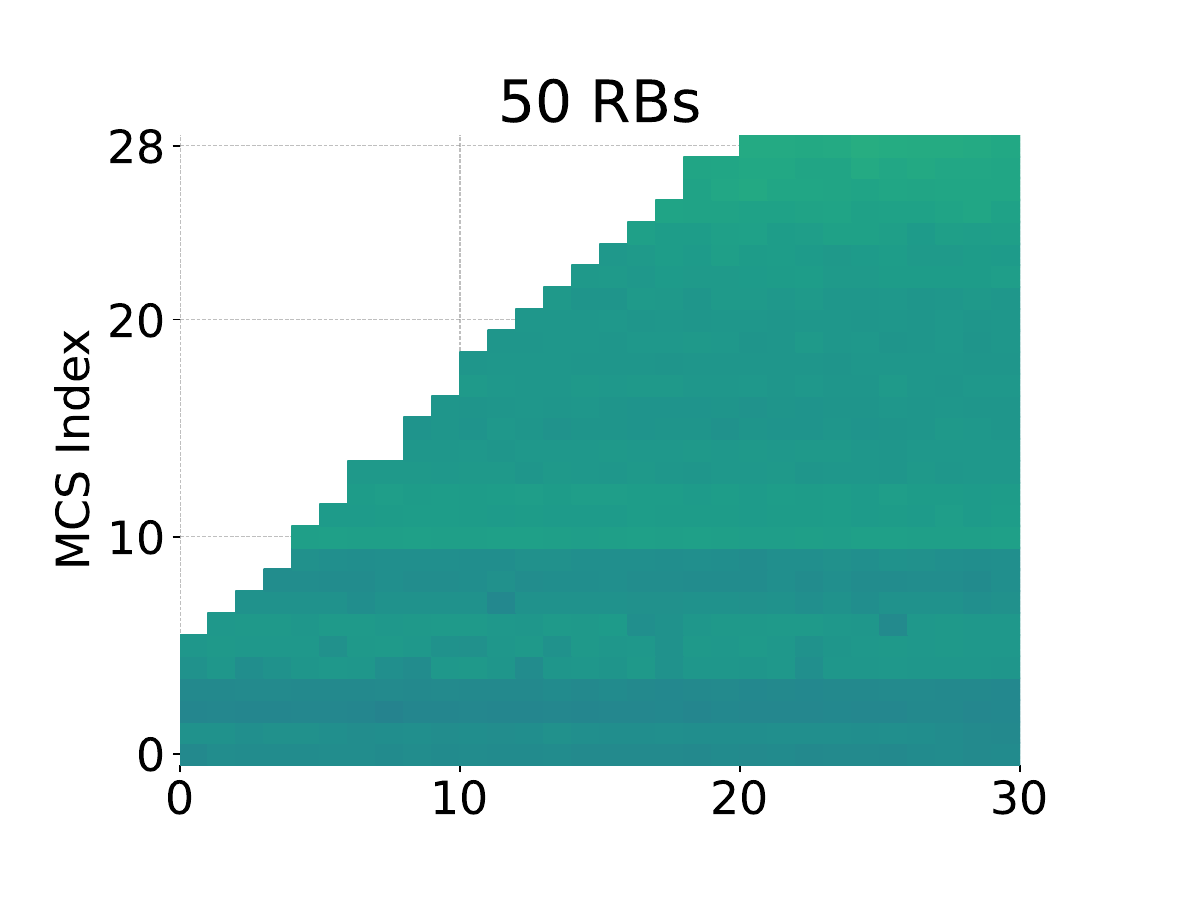}
        \label{fig:3090-50tb}}
        \hspace{-8mm}
    \subfloat{
        \includegraphics[width=0.26\textwidth]{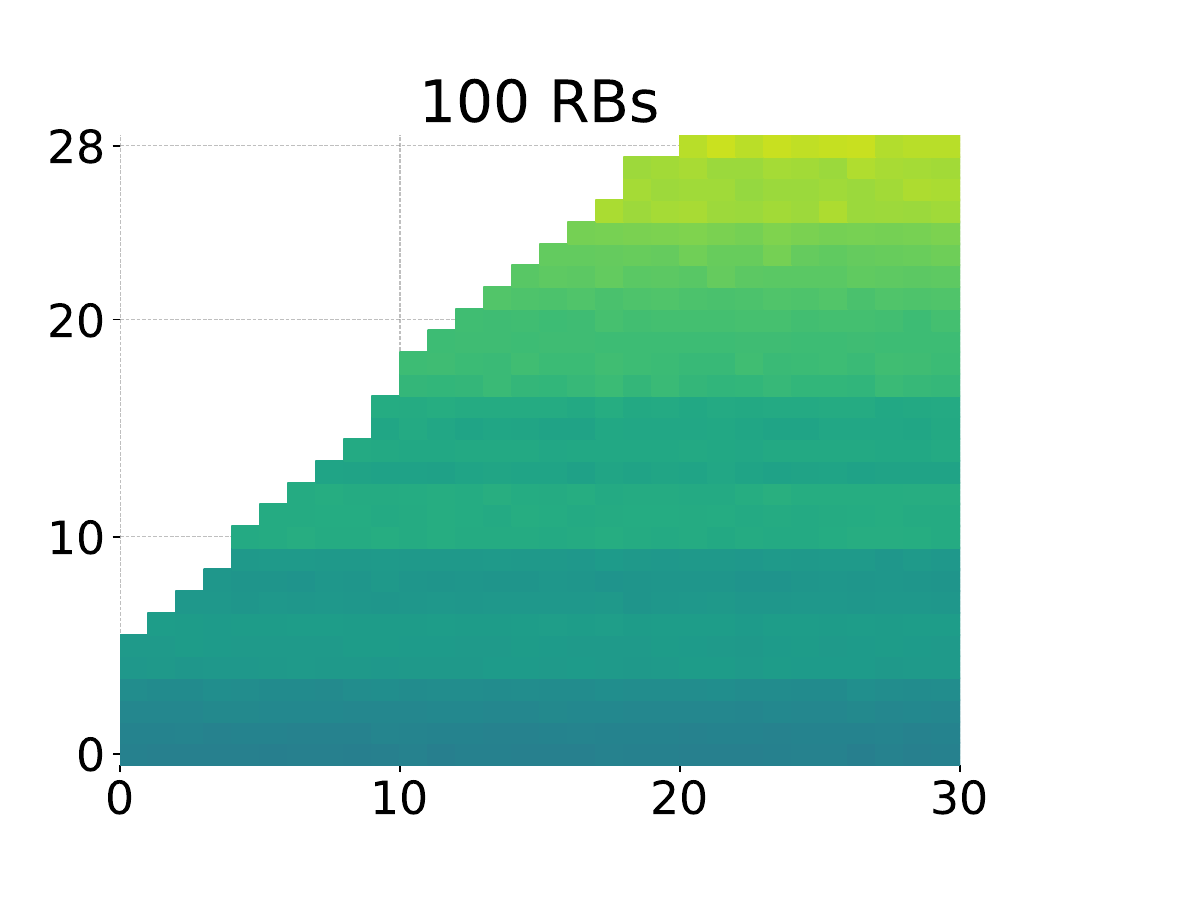}
        \label{fig:3090-100tb}}
        \hspace{-10mm}
    \subfloat{
        \includegraphics[width=0.26\textwidth]{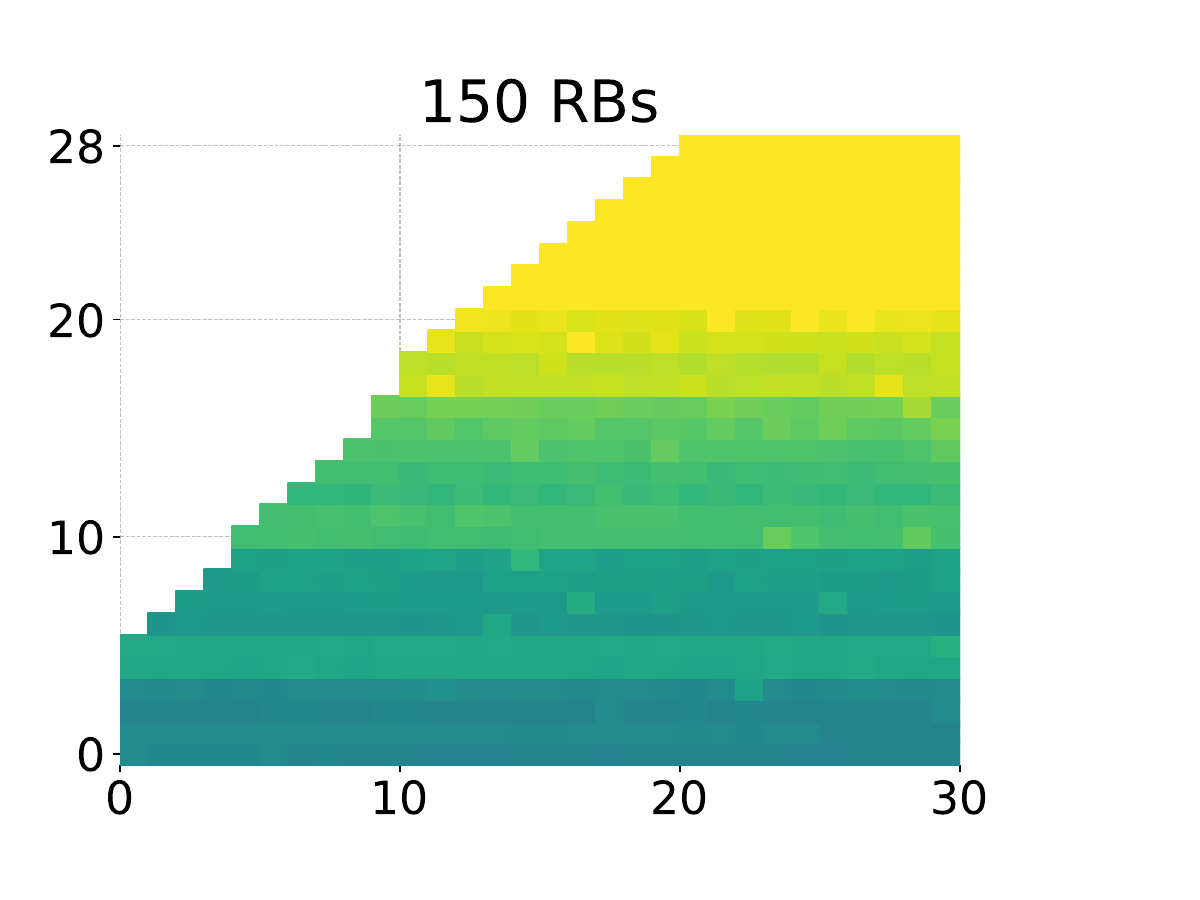}
        \label{fig:3090-150tb}}
        \hspace{-10mm}
    \subfloat{
        \includegraphics[width=0.26\textwidth]{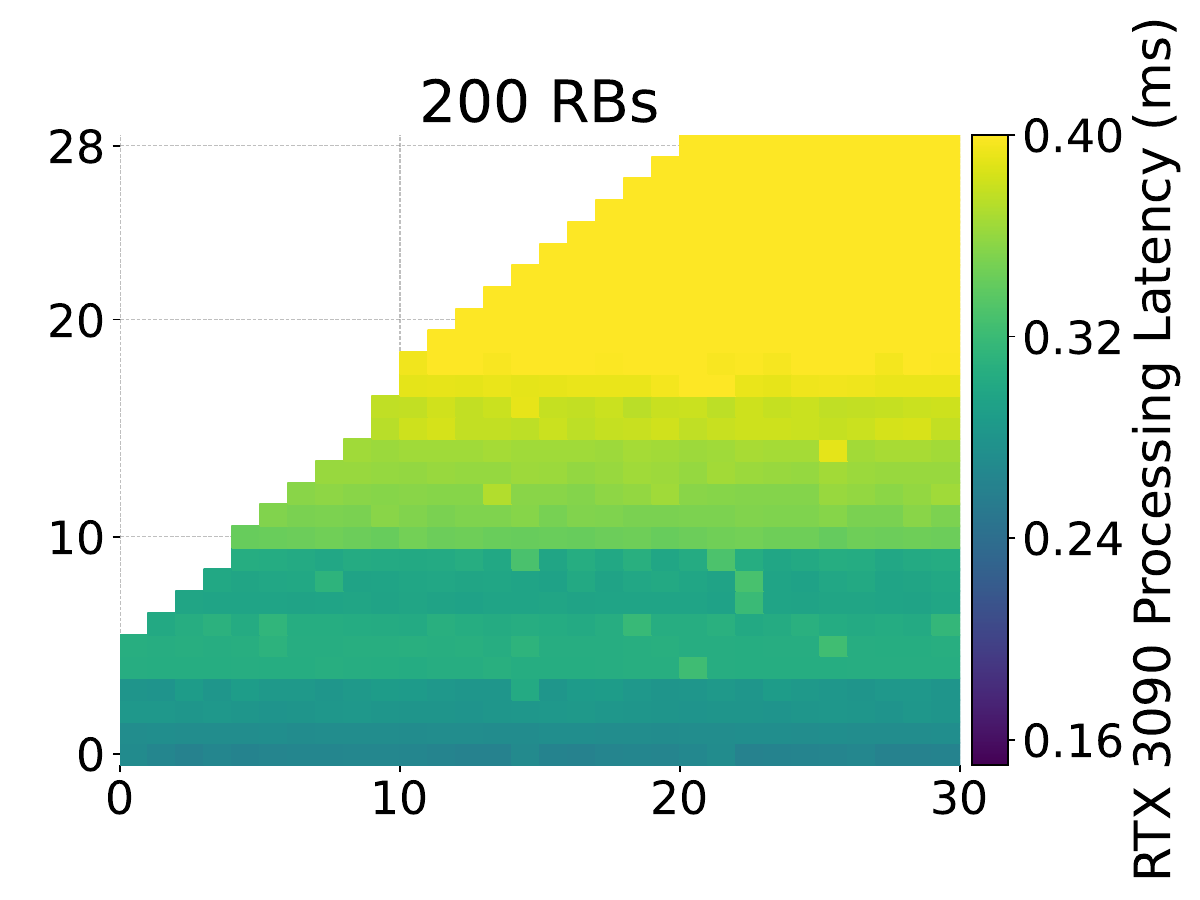}
        \label{fig:3090-200tb}} \\
    \vspace{-3mm}
    \subfloat{
        \includegraphics[width=0.26\textwidth]{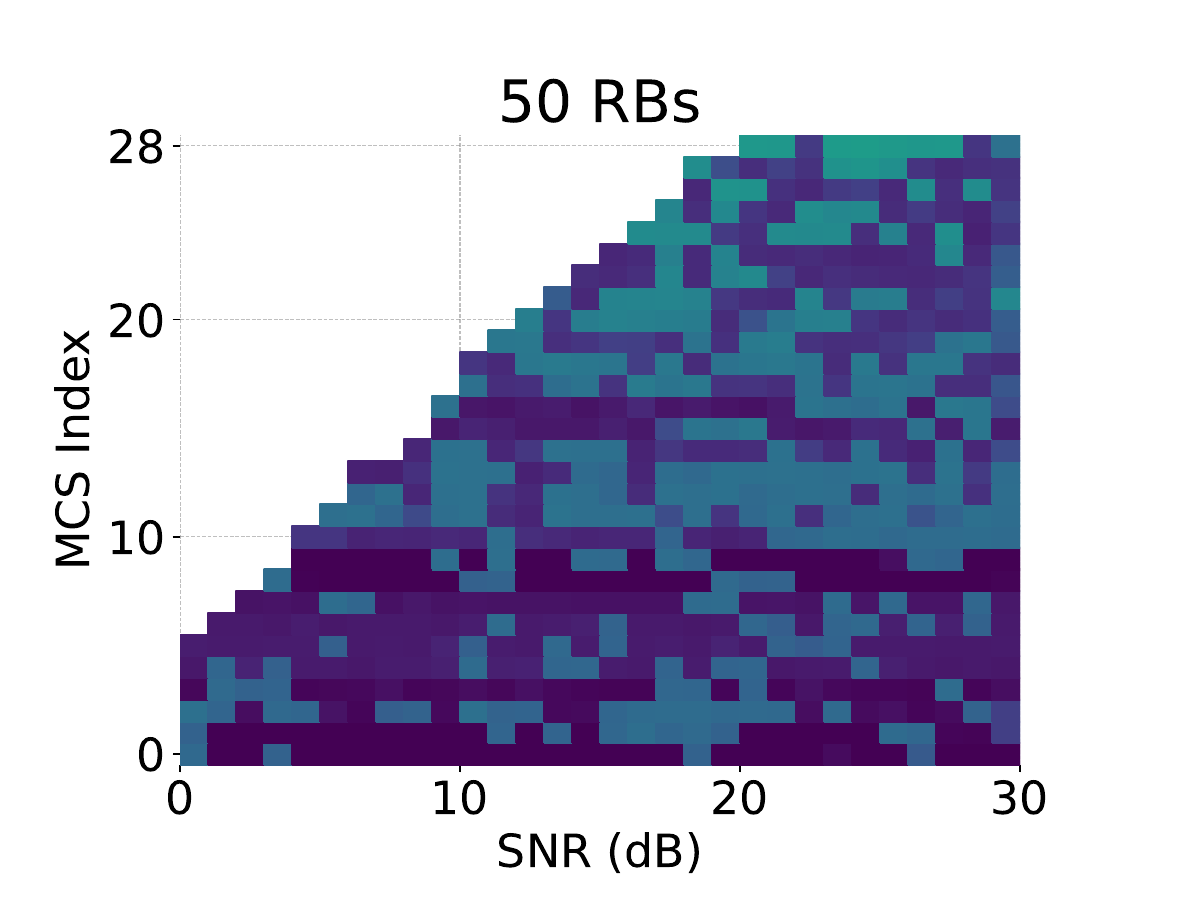}
        \label{fig:6000-50tb}}
        \hspace{-8mm}
    \subfloat{
        \includegraphics[width=0.26\textwidth]{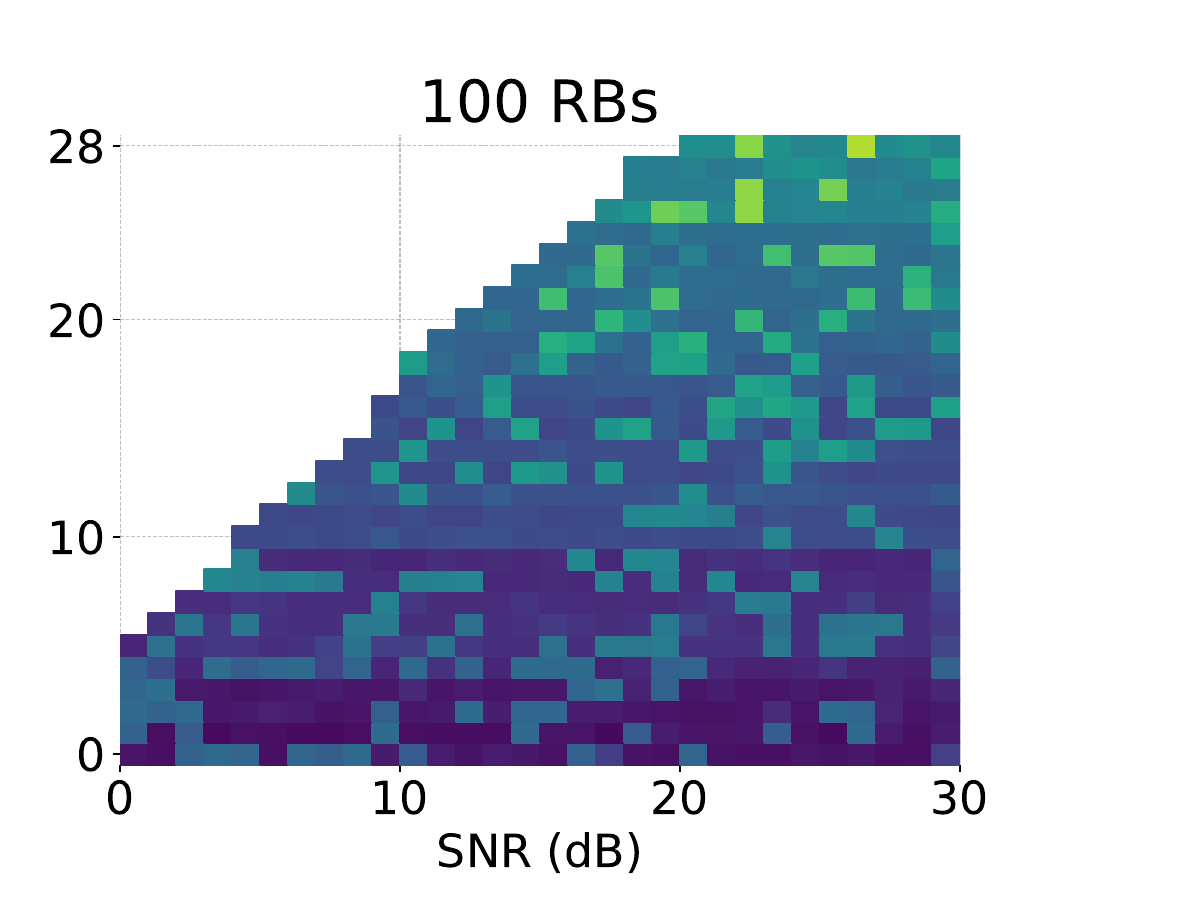}
        \label{fig:6000-100tb}}
        \hspace{-10mm}
    \subfloat{
        \includegraphics[width=0.26\textwidth]{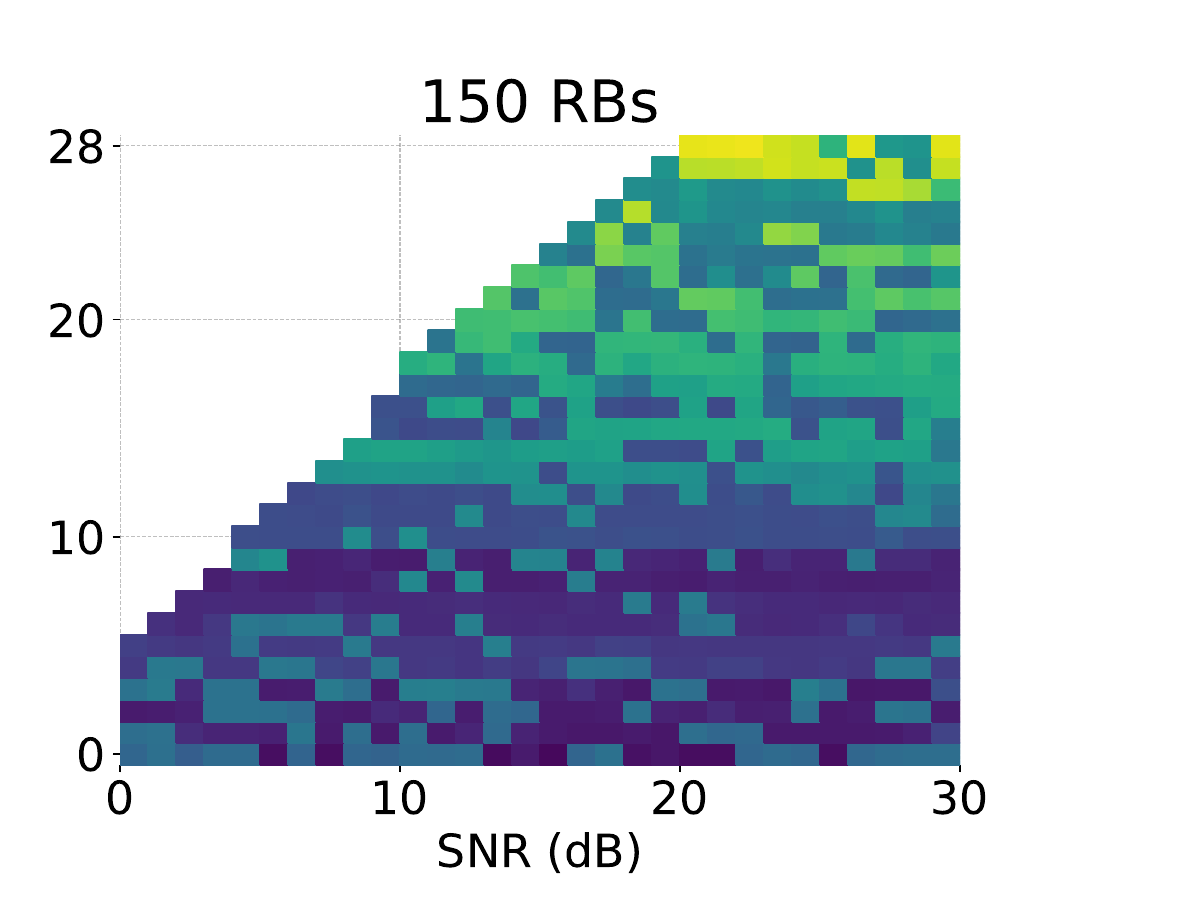}
        \label{fig:6000-150tb}}
        \hspace{-10mm}
    \subfloat{
        \includegraphics[width=0.26\textwidth]{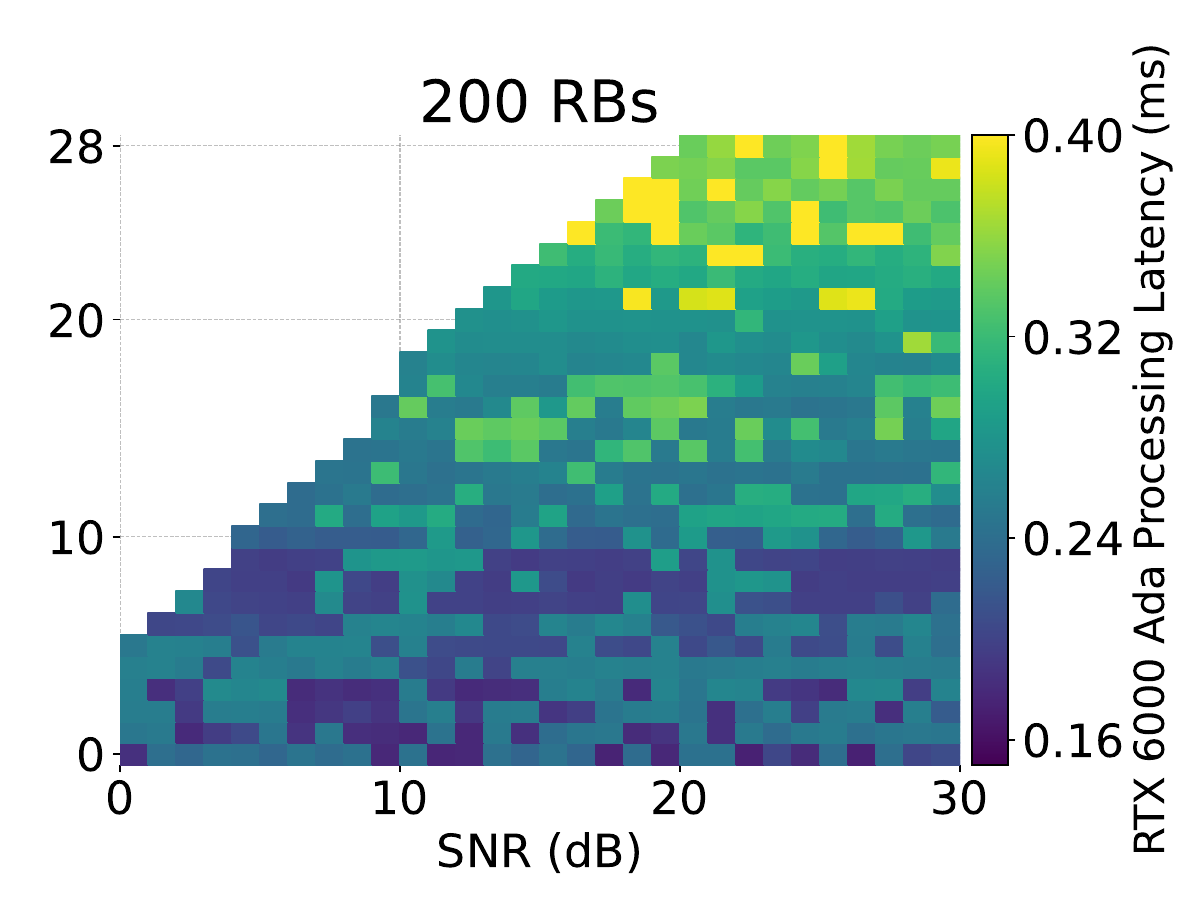}
        \label{fig:6000-200tb}}

    \vspace{-4mm}
    \caption{\small 
    LDPC decoding latency across heterogeneous platforms under varying L1 configurations.
    Each heatmap shows per-TB decoding latency (ms) as a function of MCS index and SNR level, for four PRB allocations: 
    (\emph{1st}) CPU (FlexRAN, AVX512);
    (\emph{2nd}) ACC100 lookaside acceleration; and
    (\emph{3rd}) H200 decoding via NVIDIA's Aerial. 
    (\emph{4th}) RTX 3090 decoding via NVIDIA's Aerial. 
    (\emph{5th}) RTX 6000 Ada decoding via NVIDIA's Aerial. 
    }
    \vspace{-2mm}
    \label{fig: latency}
\end{figure*}

\subsection{DecodeGPU-SionnaRK}
The NVIDIA Jetson Orin AGX~\cite{nvidia-jetson} platform represents a class of edge AI systems-on-chip (SoCs) that integrate CPU, GPU, and memory subsystems within a unified architecture. 
Unlike discrete accelerator platforms that rely on PCIe-based data transfers, Jetson adopts a \emph{unified memory} design in which the CPU and GPU share a single physical DRAM space managed by a coherent memory controller. 
This design eliminates explicit host-device data copies and allows intermediate baseband data to be accessed directly by both CPU and GPU threads. 
From a memory management perspective, it behaves as a hybrid between inline and lookaside architectures, as illustrated in the bottom panel of Fig.~\ref{fig: compute_flow}. 
While data still moves between CPU and GPU contexts, the overhead is significantly reduced compared to traditional split-memory systems since both processors operate on the same physical memory region. 
This property enables efficient acceleration for baseband processing tasks.

DecodeGPU-SionnaRK executes LDPC decoding through the \texttt{\small Sionna-RK}~\cite{sionna-rk} framework, which provides a TensorFlow- and CUDA-based implementation of the 5G NR LDPC chain optimized for embedded GPU platforms. 
By leveraging CUDA streams and unified memory, LDPC decoding can be performed \emph{hybrid} with other baseband kernels--such as FFT, channel equalization, and demodulation—without intermediate buffer copies. 
As depicted in Fig.~\ref{fig: compute_flow}(c), demodulated LLRs are written directly into unified memory and immediately consumed by the LDPC decoder kernel, forming a continuous GPU-resident data path. 

%% file: 5.evaluation.tex
\section{Evaluation}
\label{sec:evl}

\myparatight{Implementation.}
We run the FlexRAN FEC SDK on a dual-socket Intel Xeon Gold~6348 server with 112 hardware threads 
({2.6}\thinspace{GHz}, AVX512-enabled), equipped with {84}\thinspace{MB} L3 cache. 
The system runs Ubuntu~22.04 (kernel 6.8.0) with Intel oneAPI Base and HPC toolkits and the \texttt{\small icx} compiler for AVX512-optimized LDPC kernels. 
This platform serves as the CPU baseline.
For full inline acceleration, we also perform LDPC unit tests using NVIDIA Aerial (v25.2-cuBB) on a dual-socket AMD EPYC~9355 server (128 hardware threads, AVX512-enabled) equipped with four NVIDIA H200 NVL GPUs running Ubuntu~24.04 (kernel 6.8.0). 
To broaden our evaluation, we extend the analysis to additional discrete GPUs, including the NVIDIA RTX 3090 and RTX 6000 Ada, enabling cross-generation comparison of decoding latency and kernel efficiency.
The same server hosts two LDPC ASIC accelerators: one Intel ACC100 and one Silicom-branded equivalent device, both accessed through the DPDK BBDev drivers. 
We use an NVIDIA Jetson Orin AGX edge platform running Ubuntu~22.04 (kernel 5.15) with CUDA-accelerated LDPC decoding through the Sionna-rk (v1.2.1) toolkit. 
This platform executes all LDPC decoding kernels directly in unified system memory.

\myparatight{LDPC decoding latency.}
Fig.~\ref{fig: latency} summarizes the LDPC decoding latency across the CPU-, GPU-, and ACC100-based platforms for varying PRB sizes, MCS levels, and SNR values. 
We exclude Jetson AGX results from this comparison since its embedded-class CPU and GPU resources are not directly comparable to server-grade acceleration environments.
Both ACC100 and GPU-based accelerators substantially reduce decoding latency compared to CPU-only execution. 
On the CPU baseline, latency varies significantly with SNR, as poor channel conditions increase the number of required decoding iterations, whereas ACC100 and GPU implementations exhibit minimal SNR sensitivity.

Distinct scaling behaviors emerge across platforms. 
CPU decoding latency increases with PRB size, while ACC100 latency remains nearly constant from 50 to 200 PRBs, demonstrating that the accelerator’s deeply pipelined architecture maintains throughput even under higher data volumes. 
H200 GPU decoding via NVIDIA Aerial follows a similar trend, although PCIe orchestration overhead becomes more pronounced at smaller workloads. 
Across both accelerators, specific MCSs (notably MCS-10 and MCS-17) incur higher latency, as lower code rates introduce greater redundancy and decoding complexity. 
Overall, the latency profiles of ACC100 and H200 GPU align closely under sequential processing.

With other GPU models, we observe that the RTX 3090 exhibits latency behavior intermediate between the CPU and H200 platforms; its performance remains MCS-dependent and consistently higher than that of the H200. 
By contrast, the RTX 6000 Ada behaves differently from all other devices: its irregular latency pattern likely stems from architectural differences (SM89) and the lack of native PyAerial support~\cite{sm89} for this GPU generation, meaning kernel scheduling and memory handling may not align with Aerial’s optimized execution paths. 
This further underscores the value of the {\name} benchmark in identifying software–hardware mismatches and guiding future vRAN deployments across heterogeneous GPU architectures.

\begin{figure}[!t]
    \centering
    \vspace{-4mm}
    \subfloat{
        \includegraphics[width=0.47\columnwidth]{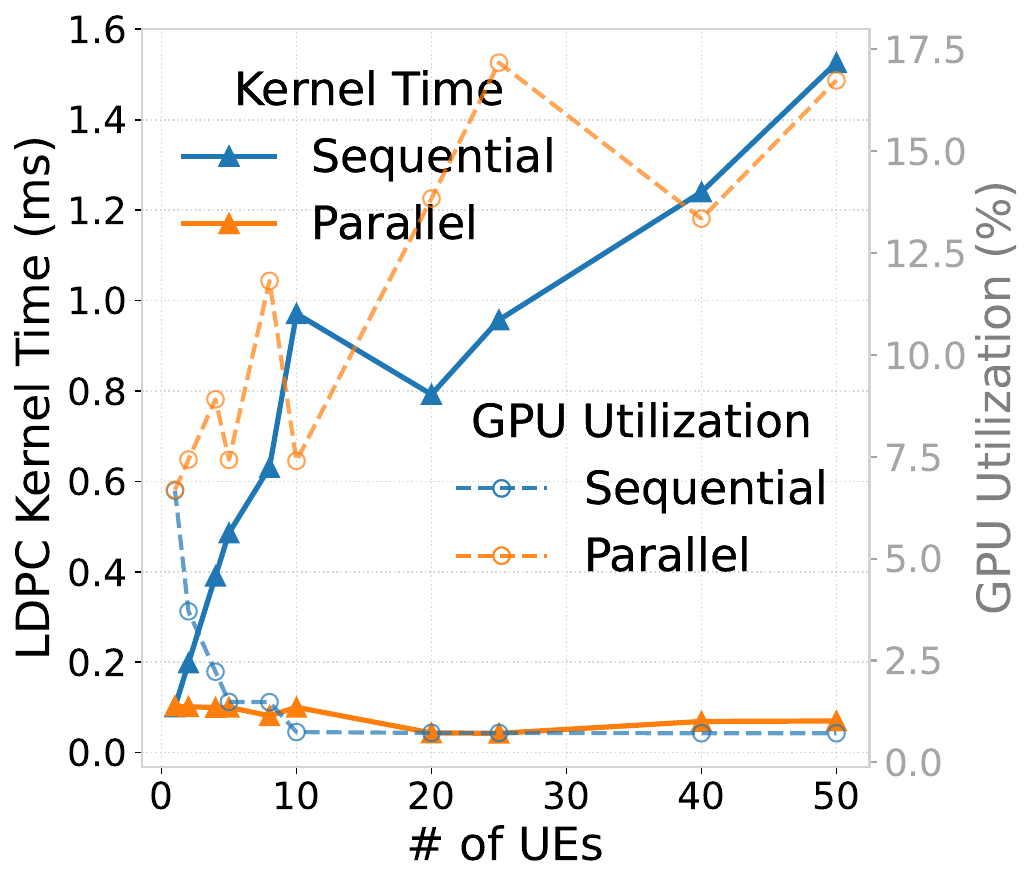}
        \label{fig:kernel-time}}
    \hspace{1mm}
    \vspace{-2mm}
    \subfloat{
        \includegraphics[width=0.45\columnwidth]{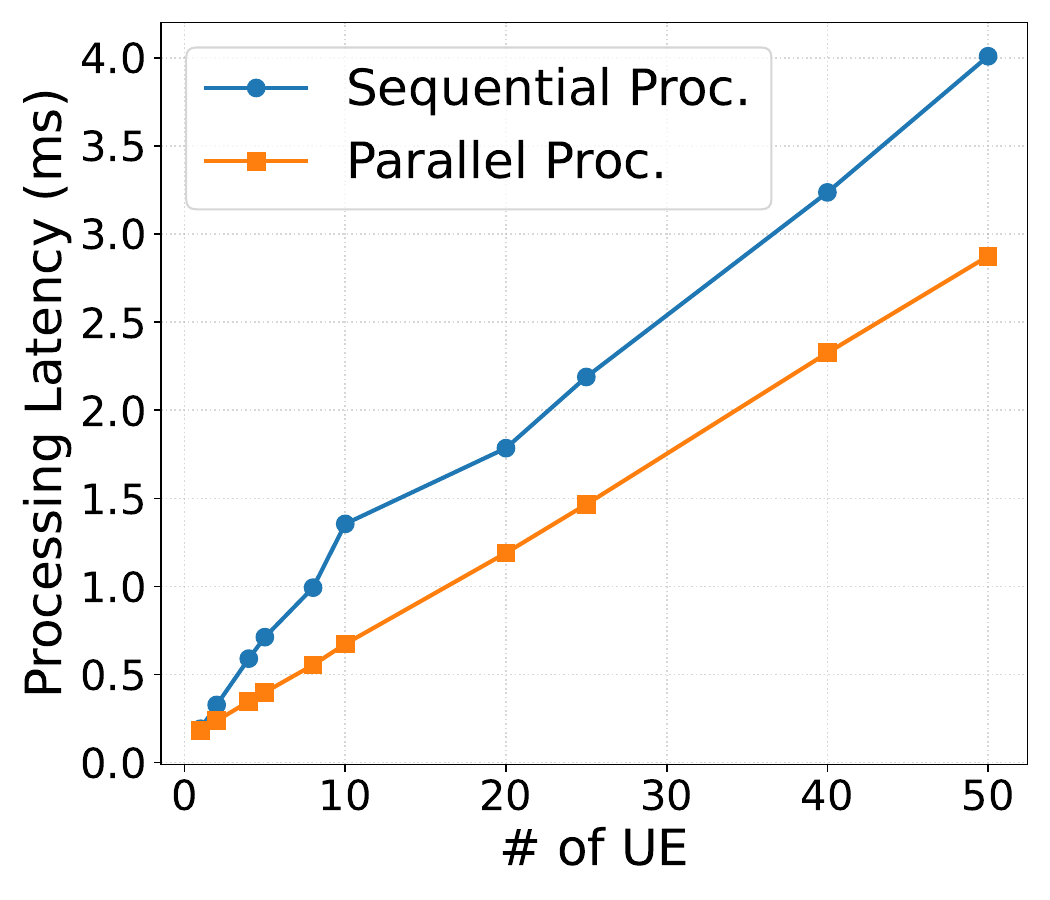}
        \label{fig:gpu-time}}
    \vspace{-2mm}
    \caption{\small PyAerial LDPC decoding performance under sequential and parallel processing. 
    (a) GPU kernel execution time and utilization, excluding host-device data movement. 
    (b) Overall LDPC decoding latency, including CPU orchestration and data transfer overheads. 
    }
    \vspace{-5mm}
    \label{fig: tp_parallel_sequential}
\end{figure}

\myparatight{Exploring GPU parallelism.}
To exploit the parallelism on GPU, Fig.~\ref{fig: tp_parallel_sequential} compares DecodeGPU-Aerial performance under sequential and parallel LDPC decoding. 
In sequential processing, the data stream for each  UE is decoded independently, one TB at a time; while parallel processing decodes $n$ TBs simultaneously across multiple CUDA grids and blocks. 
For fairness, the total data volume remains constant across both settings; for instance, with 10 UEs and 200 PRBs assigned, sequential processing assigns 20 PRBs per UE and decodes them sequentially, whereas parallel processing decodes all PRBs assigned to the 10 UEs concurrently. 
When considering only the GPU kernel execution time in Fig.~\ref{fig: tp_parallel_sequential}(a), parallel decoding achieves up to $20\times$ smaller latency than sequential decoding but with increased GPU utilization. 
Sequential processing maintains a nearly constant utilization regardless of UE count, while parallel execution scales utilization linearly with the number of active streams. 
This observation highlights that host-device data transfer and orchestration costs dominate the overall latency, reinforcing the advantage of \emph{inline processing} designs such as Aerial, where data remains resident on the GPU to minimize transfer overhead.

%% file: 6.discussion.tex
\section{Discussions}
\label{sec:diss}

\begin{figure}[!t]
    \centering
    \vspace{-3mm}
    \subfloat{
        \includegraphics[width=0.55\columnwidth]{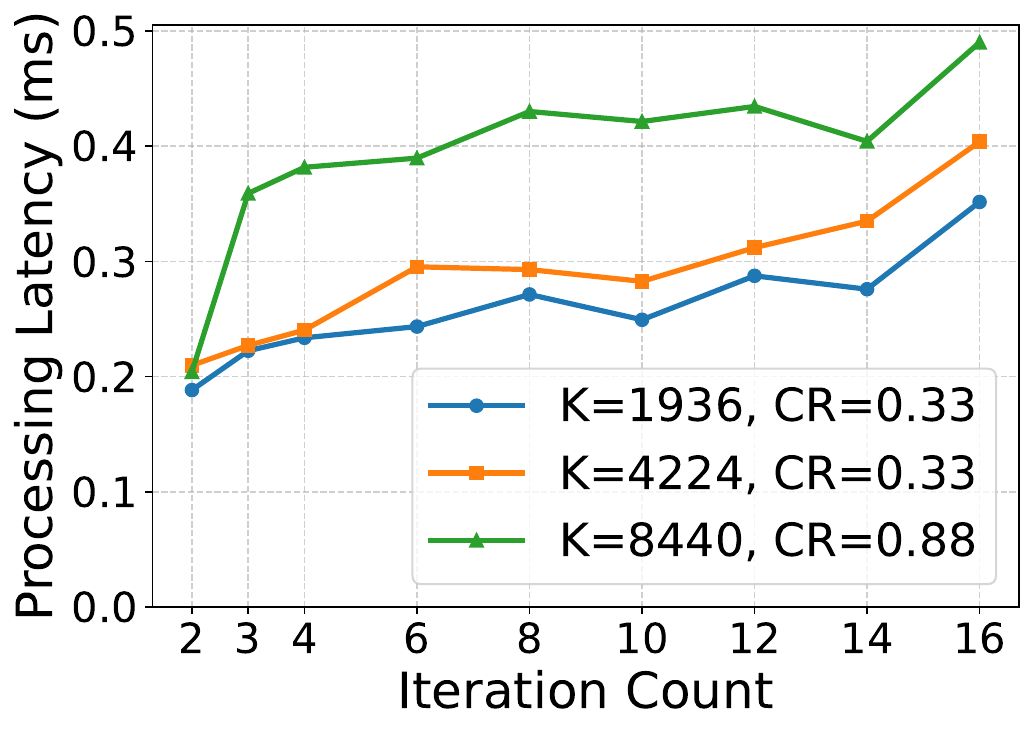}
        }
    \vspace{-2mm}
    \caption{\small LDPC decoding latency on \texttt{\small DecodeGPU-SionnaRK} across different iteration counts, with varying number of information bits ($K$) and code rates (CR).}
    \vspace{-5mm}
    \label{fig: jetson_latency}
\end{figure}

\myparatight{DecodeGPU-SionnaRK on Jetson Orin AGX.}
We also present profiling results for DecodeGPU-SionnaRK as shown in Fig.~\ref{fig: jetson_latency}. 
Since the current implementation does not yet support the full MCS set with variable code rates, we evaluate three representative cases: $K = 1,936$ and $K = 4,224$ with a code rate of $0.33$, and $K = 8,440$ with a code rate $0.88$. 
Here, $K$ denotes the number of information bits per TB. 
The overall results resemble a similar trend observed in the CPU-based decoding results (see Fig.~\ref{fig: latency}): larger transport block sizes lead to higher processing latency, while varying the amount of redundancy (code rate) has a relatively smaller impact; and the processing latency also increases with the iteration count, which corresponds to lower-SNR decoding conditions.
Overall, Jetson Orin AGX exhibits similar scaling characteristics to CPU decoding, as both the embedded GPU and CPU in this research-oriented platform have limited computational throughput compared to discrete accelerators such as the ACC100 or datacenter-grade GPUs.

\myparatight{Optimizing GPU-based L1 processing.}
Our preliminary investigation of {\namegpu} reveals that further optimization can be achieved by refining how GPU kernels and CUDA streams are organized across the L1 processing pipeline. 
Future optimization may focus on fusing adjacent kernels or reusing persistent kernels to reduce kernel bring-up latency. 
Another direction is to improve stream-level parallelism by dynamically allocating multiple CUDA streams per group of subcarriers, thereby increasing concurrency across OFDM symbols and fully exploiting the subcarrier-level parallelism inherent in the 5G NR waveform. 
Additionally, exploring more fine-grained task scheduling within the GPU’s streaming multiprocessors could help maintain high utilization under variable TB sizes and traffic conditions. 

\myparatight{Energy-aware and dynamic compute resource allocation for vRANs.}
Allocation of heterogeneous compute resources must account not only for RAN throughput and latency, but also for the energy consumed by computation and data movement, where the latter often dominates under low compute utilization.
Our recent work \emph{NEXUS}~\cite{qi2025nexus} explored dynamic and energy-aware baseband scheduling across CPU and ACC100 accelerators, demonstrating that joint optimization of computation and data-transfer pipelines can significantly improve energy proportionality in heterogeneous vRAN systems.
Complementary to this direction, prior work~\cite{annamalai2025exploring} investigates adaptive resource provisioning in large-scale radio deployments for balancing latency and energy efficiency. 
Unlike CPUs or dedicated accelerators, GPUs exhibit highly dynamic power behavior that depends on kernel concurrency, memory access intensity, and streaming multiprocessor utilization. 
Future work may explore regression- or learning-based models that map kernel configurations (e.g., MCS, iteration count) to power consumption to enable adaptive and energy-aware scheduling.

\myparatight{Coexistence with ML workloads.}
A frequently raised concern pertains to the high capital cost and suboptimal energy efficiency of GPUs when dedicated solely to baseband processing. 
However, such concerns can be mitigated if the same hardware is shared dynamically with other compute-intensive tasks when the traffic load is low. 
Recent work~\cite{foukas2025future} also advocates for a converged AI-cellular architecture, motivating the co-location of communication and computation workloads to support industrial-scale AI applications. 
For instance, during non-URLLC periods or under light traffic conditions, GPU resources used for L1 tasks can be time- or spatially-partitioned to support large ML model training and/or inference workloads. 
This flexible multi-tenancy model highlights the potential of GPU-based architectures to unify communication, sensing~\cite{fang2022joint, hunt2023madradar}, and computation workloads within in the vRAN context.

%% file: 7.conclusion.tex
\section{Conclusions}
\label{sec:conclusions}

We presented {\name}, a unified benchmarking framework that systematically evaluates LDPC decoding performance across heterogeneous computing platforms, including CPUs, GPUs, and ASICs. 
Through careful implementation and detailed profiling, we highlighted how different system architectures organize their L1 processing pipelines, and how factors such as kernel launch overhead, data-movement orchestration shape overall decoding latency and determinism. 
As an open and extensible benchmark, {\name} provides a foundation for future research on cross-platform baseband processing, facilitating energy-aware compute resource management for heterogeneous vRAN systems.